\documentclass[journal]{IEEEtran}

\usepackage{pdfpages} 

\usepackage{setspace}
\ifCLASSINFOpdf
\else
\fi
\usepackage[cmex10]{amsmath}

 \usepackage[hyperindex=true,pdftitle={},
pdfauthor={St\'ephane Guerrier},colorlinks=TRUE,
pagebackref=false,citecolor=blue,plainpages=false,
pdfpagelabels]{hyperref} 

\usepackage{verbatim} 
\usepackage{amsthm}
\usepackage{amsbsy}
\usepackage{amssymb}
\usepackage{amsbsy}
\usepackage{amsfonts}
\usepackage{graphics}
\usepackage{bm}
\usepackage{bbm}
\usepackage{cite}
\usepackage{mathtools}
\usepackage{authblk}
\usepackage{booktabs}
\usepackage{nicefrac}
\usepackage{enumitem}
\usepackage{bbding}
\usepackage{soul}
\usepackage[numbers]{natbib}
\usepackage{nicefrac}
\usepackage{placeins}

\usepackage{xr-hyper} 
\DeclareMathOperator*{\argmin}{argmin}
\DeclareMathOperator*{\cov}{cov}

\DeclareSymbolFont{lettersA}{U}{txmia}{m}{it}
\DeclareMathSymbol{\real}{\mathord}{lettersA}{"92}
\DeclareMathSymbol{\field}{\mathord}{lettersA}{"83}

\DeclarePairedDelimiter\floor{\lfloor}{\rfloor} 

\def\boxit#1{\vbox{\hrule\hbox{\vrule\kern3pt
          \vbox{\kern3pt#1\kern3pt}\kern3pt\vrule}\hrule}}

\def\sgcomment#1{\vskip0mm\boxit{\vskip 0mm{\color{blue}\bf#1}
     {\color{blue}\bf -- SG\vskip 1mm}}\vskip 0mm}

\def\hxcomment#1{\vskip0mm\boxit{\vskip 0mm{\color{green}\bf#1}
     {\color{blue}\bf -- HX\vskip 1mm}}\vskip 0mm}
\definecolor{pinegreen}{rgb}{0.0, 0.47, 0.44} 
  
\newtheoremstyle{mytheoremstyle} 
    {0.3cm}                      
    {0cm}                        
    {\itshape}                   
    {}                           
    {\scshape}                   
    {: }                          
    {0em}                       
    {}  

\theoremstyle{mytheoremstyle}
\newtheorem{Theorem}{Theorem}
\newtheorem{Lemma}{Lemma}
\newtheorem{Proposition}{Proposition}
\newtheorem{Corollary}{Corollary}

\renewenvironment{proof}{{\noindent \sc Proof:}}{\qed}

\newtheoremstyle{myExampleRemarkstyle} 
    {0.3cm}                    
    {0cm}                           
    {\itshape}                   
    {}                           
    {\scshape}                   
    {: }                          
    {0em}                       
    {}  

\theoremstyle{myExampleRemarkstyle}

\newtheoremstyle{simuStyle}
{0.3cm} 
{0cm} 
{} 
{} 
{\bfseries} 
{.} 
{0em} 
{} 

\theoremstyle{simuStyle}

\newtheoremstyle{stratStyle}
{0.3cm} 
{0cm} 
{} 
{} 
{\scshape} 
{: } 
{0em} 
{} 

\theoremstyle{stratStyle}

\DeclareSymbolFont{lettersA}{U}{txmia}{m}{it}
\DeclareMathSymbol{\real}{\mathord}{lettersA}{"92}
\DeclareMathSymbol{\field}{\mathord}{lettersA}{"83}

\usepackage[final]{changes}
\newif\iffoo

\definechangesauthor[name=Reviewer 1, color = purple]{r1}
\definechangesauthor[name=Reviewer 2, color = orange]{r2}
\colorlet{Changes@Color}{red}

\usepackage[boxed]{algorithm2e}

\begin{document}

\title{Multivariate Signal Modelling\\with Applications to Inertial Sensor Calibration}
%
%
%
\author{Haotian Xu, St\'{e}phane Guerrier,
     Roberto Molinari \& Mucyo Karemera 
\thanks{\textbf{H. Xu} is a PhD student, Geneva School of Economics and Management, University of Geneva, Switzerland, 1205, Switzerland. (E-mail: haotian.xu@unige.ch).}%
\thanks{\textbf{S. Guerrier} is an Assistant Professor, Geneva School of Economics and Management, University of Geneva, Switzerland, 1205, Switzerland. (E-mail: stephane.guerrier@unige.ch).}%
\thanks{\textbf{R. Molinari} is a Postdoctoral Scholar, Department of Statistics, Pennsylvania State University, PA, 16801, USA. (E-mail: rum415@psu.edu)}
\thanks{\textbf{M. Karemera} is a Postdoctoral Scholar, Geneva School of Economics and Management, University of Geneva, Switzerland, 1205, Switzerland. (E-mail: mucyo.karemera@unige.ch)}}


%
%

\markboth{}%
{Shell \MakeLowercase{\textit{et al.}}: Bare Demo of IEEEtran.cls for Journals}
%




\maketitle

\begin{abstract}
The common approach to inertial sensor calibration for navigation purposes has been to model the stochastic error signals of individual sensors independently, whether as components of a single inertial measurement unit (IMU) in different directions or arrayed in the same direction for redundancy. These signals usually have an extremely complex spectral structure that is often described using latent (or composite) models composed by a sum of underlying models which are challenging to estimate. A large amount of research in this domain has been focused on the latter aspect through the proposal of various methods that have been able to improve the estimation of these models both from a computational and a statistical point of view. However, albeit challenging, the separate calibration of the individual sensors is still unable to take into account the dependence between each of them which can have an important impact on the precision of the navigation systems. In this paper we develop a new approach to simultaneously model both the individual signals as well as the dependence between them by studying the quantity called Wavelet Cross-Covariance and using it to extend the application of the Generalized Method of Wavelet Moments to this setting. This new method can be used in many other settings for multivariate time series modelling, especially in cases where the dependence among signals may be hard to detect since it can be based on a shared underlying model that has a marginal contribution to the processes' overall variance. Moreover, in the field of inertial sensor calibration, this approach can deliver important contributions among which the possibility to test dependence between sensors, integrate their dependence within the navigation filter and construct an optimal virtual sensor that can be used to simplify and improve navigation accuracy. The advantages of this method and its usefulness for inertial sensor calibration are highlighted through a simulation study and an applied example with a small array of XSens MTi-G IMUs.

\end{abstract}



\begin{IEEEkeywords}
Generalized Method of Wavelet Moments, Wavelet Covariance, Multivariate Time Series, Signal Processing, Virtual Sensors, IMU, Navigation
\end{IEEEkeywords}

\section{Introduction}
\label{sec:intro}

The modelling of multivariate time series is an important as well as challenging task in many applied domains of data analysis. Indeed, it is extremely common to measure phenomena over time which not only manifest autocorrelation with their past but also show a form of dependence between them. This setting can be observed, among others, in a variety of longitudinal studies \replaced[id = r1]{\cite{roy2000latent,cho2016analysis}}{\citep[see][]{roy2000latent,cho2016analysis}}, in economic and financial research \replaced[id = r1]{\cite{campbell1997econometrics, sims1980macroeconomics,koopman2000fast,pan2008modelling}}{\citep[see][]{campbell1997econometrics, sims1980macroeconomics,koopman2000fast,pan2008modelling}} as well as in different fields of engineering \replaced[id = r1]{\cite{donoho2006compressed,vaccaro2017reduced}}{\citep[see][]{donoho2006compressed,vaccaro2017reduced}}. In the latter case, an increased attention has been given to the modelling of multivariate signals for the task of inertial sensor calibration where the error signals of the individual accelerometers and gyroscopes that compose the Inertial Measurement Units (IMU) are characterized by error signals that are dependent on each other \added[id = r1]{\cite{vaccaro2017reduced}}. To date, these error signals have been modeled independently from each other and this task alone has already been challenging to deal with. In fact, the individual error signals of these instruments are characterized by deterministic and stochastic components where the first can be dealt with through physical models (which are employed for calibration and compensation prior to use) while the latter are often described through complex stochastic models that are made by the sum of different underlying processes. The estimation of these stochastic models has always been complicated and many methods used for this purpose have either suffered from statistical limits and/or are computationally or numerically unstable. Indeed, in most latent variable modelling scenarios, the Maximum-Likelihood Estimator (MLE) cannot be directly applied since the latent stochastic components are unobservable and their marginal distribution is usually unknown. A feasible likelihood-based method is the Expectation-Maximization (EM) algorithm \replaced[id = r1]{\cite{dempster1977maximum}}{\citep[see][]{dempster1977maximum}} whose implementation is nevertheless limited due to the complexity and, more importantly, the possible non-convexity of its associated optimization problems therefore severely limiting the use of its statistical properties in practice \replaced[id = r1]{\cite{anandkumar2012method}}{\citep[see][]{anandkumar2012method}}. To avoid the problems of the likelihood-based methods, the linear regression based on the Allan Variance (AVLR) is widely used for estimation and prediction especially in the field of inertial sensor calibration \replaced[id = r1]{\cite{el2008analysis}}{\citep[see][]{el2008analysis}}. Although, the AVLR is a computationally feasible method for estimating many complex models, its statistical properties are not optimal as discussed in \cite{guerrier2016theoretical} where the inconsistency of this method was shown for the majority of the latent models used for the purpose of sensor calibration.

Given the complexity of these models and the difficulties in estimating them, the signals have thus far been dealt with individually without taking into account the forms of dependence between them. As mentioned earlier, this approach is non-optimal since a multivariate approach to signal analysis would be more appropriate for different reasons. Firstly, it is reasonable to assume that there exists a dependence between individual sensors along the three axes of an IMU and, when using an array of sensors, it is even more reasonable to assume that there is dependence between error signals of sensors placed in the same direction since they are measuring the same phenomenon \added[id = r1]{\cite{bayard2002foundations}}. If the multivariate dependence is not considered, this can deliver different problems which include parameter misestimation and flawed testing procedures leading to incorrect modelling and degraded navigation performance. Considering this, the literature in the area of multivariate time series analysis has collected a variety of models to describe these settings as well as proposing methods that can estimate them. One of the first models to be proposed was the Vector Auto-Regression (VAR) model which, in some sense, generalizes the Auto-Regression (AR) model in order to allow lagged observations from some signals to influence current observations of others \added[id = r1]{\cite{hamilton1994time}}. This class of models has been extended to the more general class of Vector Auto-Regression Moving Average (VARMA) models which also allow for correlation between innovation processes in each time series \replaced[id = r1]{\cite{hamilton1994time}}{\citep[see][]{hamilton1994time}}. Although these models are quite flexible in considering different forms of autocorrelation and correlation between time series, their estimation can sometimes be numerically challenging and furthermore they may not be able to capture more complicated forms of multivariate dependence \replaced[id = r1]{\cite{lutkepohl2006forecasting,metaxoglou2007maximum}}{\citep[see][]{lutkepohl2006forecasting,metaxoglou2007maximum}}. Moreover, in practice, VARMA models are only used directly when the number of signals composing the multivariate time series is smaller than three since these models are often overparametrized thereby leading to identifiability issues \added[id = r1]{\cite{chang2018principal}}. In this case, a more appropriate approach could be represented by only considering a model (or some models) to describe the dependence between all or some subsets of the signals composing the multivariate time series. Indeed, the individual time series can often be characterized by a latent model structure in which different underlying processes are superposed and, within this setting, there can also be a dependence of these individual latent processes with those present in other signals \replaced[id = r1]{\cite{guerrier2013wavelet,harvey1990forecasting,koopman2000fast}}{\citep[see][]{guerrier2013wavelet,harvey1990forecasting,koopman2000fast}}. We refer to these models as multivariate latent models which are particularly \replaced[id = r1]{relevant}{pertinent} for the task of inertial sensor calibration.

Considering the above setting, this paper proposes a new method that is able to estimate the mentioned multivariate latent models in a computationally efficient and numerically stable manner. This new method consists in an extension of the Generalized Method of Wavelet Moments (GMWM) which was initially proposed as a statistically consistent approach to estimating complex latent models in univariate settings, overcoming the theoretical and computational limitations of the existing methods mentioned earlier \replaced[id = r1]{\cite{guerrier2013wavelet}}{\citep[see][]{guerrier2013wavelet}}. To do so the GMWM takes advantage of a quantity called the Wavelet Variance (WV) which is the variance of the wavelet coefficients \replaced[id = r1]{resulting from}{issued from} a wavelet decomposition of a signal \replaced[id = r1]{\cite{percival1995estimation}}{\citep[see][]{percival1995estimation}}. However, the WV does not take into account the variability in the signal that is explained by the dependence on other signals. For this reason, in this paper we will make use of the Wavelet Cross-Covariance (WCCV), which was proposed \replaced[id = r1]{in \cite{whitcher2000wavelet}}{by \citet[]{whitcher2000wavelet}} and whose asymptotic properties we develop further in this paper by reducing and weakening the relative conditions. Based on the latter properties, we extend the GMWM to include this quantity and deliver the Mulitvariate GMWM (MGMWM) whose statistical properties are studied in order to perform the required inference procedures. In order to achieve the latter properties, we will also discuss the identifiability of a wide class of latent multivariate time series models which is essential to obtain consistency of the MGMWM, thereby shedding light on what kind of latent multivariate models can be estimated and reducing the conditions necessary to deliver adequate inference procedures on them. Based on these properties, the proposed framework allows to model dependence between and within sensors as well as provide the tools to test the presence of multivariate dependence between stochastic error signals thereby facilitating sensor calibration and contributing to navigation accuracy.

This paper is organized as follows. In Section \ref{sec:multproc} we introduce the class of multivariate latent processes that we are going to propose and study, justifying their relevance for the task of inertial sensor calibration. In Section \ref{sec:wccv} we briefly introduce the wavelet decomposition to then define and study the WCCV thereby delivering the asymptotic properties of the WCCV vector. The latter properties are essential, among others, in order to obtain the asymptotic properties of the MGMWM which is introduced in Section \ref{sec:mgmwm} where the identifiability of a large class multivariate latent models is studied. To highlight the good statistical properties of the new estimator as well as its usefulness, Section \ref{sec:simulation} presents a simulation study comparing the finite sample performance of the proposed estimator with a recently proposed alternative while Section \ref{sec:applications} presents an application in the field of inertial sensor calibration where this new approach can deliver an important contribution. Finally, Section \ref{sec:conclusions} concludes.

\section{Multivariate Latent Processes}
\label{sec:multproc}

Let us define a multivariate process as $(\mathbf{X}_t )_{t \in \mathbb{Z}}\subset\mathbb{R}^{I}$, where $I \in \mathbb{N}$. As mentioned in the introduction, one of the most common approaches to modelling these multivariate process is through the use of VARMA models which have the following structure:
\begin{equation*}
\mathbf{X}_t \equiv \sum_{p=1}^P \mathbf{A}_p \mathbf{X}_{t-p} + \sum_{q=1}^Q \mathbf{U}_q \bm{\epsilon}_{t-q} + \bm{\epsilon}_{t}  ,
\end{equation*}
where $P$ and $Q$ are the maximum lags of dependence for the AR and MA processes respectively among the individual time series composing the \replaced[id = r2]{multivariate}{mutlivariate} process, $\mathbf{A}_p$ and $\mathbf{U}_q$ are the coefficient matrices for the $p^{th}$ and $q^{th}$ lag and 
\begin{equation}
  \bm{\epsilon}_t \overset{\text{i.i.d.}}{\sim} \mathcal{F}(\mathbf{0}, \mathbf{\Sigma}) ,
  \label{eq.innov}
\end{equation}
are independently distributed innovation vectors with multivariate distribution $\mathcal{F}$, expectation $\mathbf{0}$ and covariance matrix $\mathbf{\Sigma}$.

These multivariate time series models are very useful to describe a wide variety of phenomena and, at an individual level, can often be represented as a latent process made by the sum of underlying first-order autoregressive and white noise models \added[id = r1]{\cite{granger1976time}}. However, these models hide the latent structure which can often be extremely useful for interpretation and, more importantly, are limited to a class of latent models in which many processes that are relevant to many domains cannot be included. For this reason, in this paper we consider a different class of multivariate latent models (which can possibly include reparametrizations of the aforementioned VARMA models) in order for practitioners to \replaced[id = r1]{use these}{make use of them} in a variety of settings where they are of great importance (e.g. inertial sensor calibration). To do so, let us define a multivariate time series \replaced[id = r1]{composed by $K$ latent processes}{composed of K latent processes} as
\begin{equation}
\mathbf{X}_t \equiv \sum_{k=1}^K \mathbf{S}_k \mathbf{X}_{k,t},
\label{MLP}
\end{equation}
where $\mathbf{X}_t \equiv [X_t^{(i)}]_{i=1,\hdots,I}$ is an $I$-dimensional vector and $\mathbf{X}_{k,t} \equiv [X_{k,t}^{(j)}]_{j=1,\hdots,D_k}$ is a $D_k$-dimensional vector with $k$ being the index for the type of univariate/multivariate model underlying some or all of the latent processes and $\mathbf{S}_k$ is an $I \times D_k$ matrix with either $1$ or $0$ as its elements that we define as 
\begin{equation}
(\mathbf{S}_k)_{i,j} =  
   \left\{
  \begin{array}{ll}
    1  & \mbox{if } X_{k,t}^{(j)} \mbox{ appears in } X_t^{(i)} \\
    0 & \mbox{if } \ \text{otherwise}.
  \end{array}
\right . \nonumber
\end{equation}
%
Therefore, we have a set of $K$ multivariate time series models (which can characterize one, some or all of the univariate processes) that can be summed to compose a multivariate latent model.  Given this, \replaced[id = r1]{let us introduce the individual multivariate time series models, commonly used also for the task of inertial sensor calibration (see e.g. \cite{titterton2004strapdown}), that will be considered in this paper to build a general multivariate latent model:}{let us now introduce the processes that are each characterized by one of the $K$ multivariate models that we consider in the context of this paper and that constitute the basic models used for inertial sensor calibration}

\begin{enumerate}[label=\bfseries (T\arabic*)]
    \item White Noise (WN) which corresponds to the independently distributed innovation vectors $\bm{\epsilon}_t$ defined in (\ref{eq.innov}). We denote the process characterized by this model as
    $\mathbf{X}_{1, t}$, hence $\bm{X}_{1, t} \overset{\text{i.i.d.}}{\sim} \mathcal{F}(\mathbf{0}, \bm{\Sigma})$. \label{WN}
    \item Random Walk (RW) which we denote as $\mathbf{X}_{2, t}$ and is defined as
        \begin{equation*}
        \mathbf{X}_{2, t} = 
        \mathbf{X}_{2, t-1}
        +
        \bm{\iota}_t,
    \end{equation*}
    where $\bm{\iota}_t \overset{\text{i.i.d.}}{\sim} \mathcal{F}(\mathbf{0}, \bm{\Lambda})$. \label{RW}
    \item Quantization Noise (QN) (or rounding error, see e.g. \replaced[id = r1]{\cite{papoulis2002probability}}{\citet{papoulis2002probability}}) which we denote as $\mathbf{X}_{3, t}$. For this process we do not consider a multivariate model and therefore each univariate process is characterized by the parameter $Q_i^2 \in \{x \in \mathbb{R} | x > 0\}$.  \label{QN}

    \item Drift (DR) which we denote as $\mathbf{X}_{4, t}$ and, for each univariate process, is defined as
$$X_{4,t}^{(i)} = \omega_i t,$$    
     where $\omega_i \in \{x \in \mathbb{R} | x > 0\}$.\label{DR}
    \item First-Order Auto-Regressive Noise (AR1) which we denote as $\mathbf{X}_{k, t}$, $\forall \,\, k = 5, \hdots, K$, where $K \in \mathbb{N}$, and defined as 
    \begin{equation*}
        \mathbf{X}_{k, t}
        =
        \bm{\Phi}
        \mathbf{X}_{k, t-1}
        +
        \bm{\varepsilon}_t,
    \end{equation*}
    where $\bm{\Phi}$ is a diagonal matrix with diagonal elements $\phi_k^{(i)}$, $0 < |\phi_k^{(i)}| < 1$, for $i = 1, \dots, D_k$ and $\bm{\varepsilon}_t \overset{\text{i.i.d.}}{\sim} \mathcal{F}(\mathbf{0}, \bm{Z}_k)$.  \label{AR}
\end{enumerate}
The covariance matrices $\bm{\Sigma}$, $\bm{\Lambda}$ and $\bm{Z}_k$ defined above are positive definite $D_k \times D_k$ matrices with $k = 1, 2, 5, ..., K$ \replaced[id = r2]{respectively}{repsectively}, with their respective elements being $[\sigma^{(i,i^{\prime})}]_{1 \leq i \leq i^{\prime} \leq D_1}$, $[\lambda^{(i,i^{\prime})}]_{1 \leq i \leq i^{\prime} \leq D_2}$ and $[z_k^{(i,i^{\prime})}]_{1 \leq i \leq i^{\prime} \leq D_k}$. As can be seen from the above definitions, processes \ref{QN} and \ref{DR} \replaced[id = r2]{do not}{don't} have a multivariate structure (i.e. there is no dependence between univariate signals based on these models). This is a reasonable assumption in practice since process \ref{QN} can be interpreted as a form of rounding error which is intuitively independent from the rounding error made on another signal while the \ref{DR} process is a non-stochastic process whose behavior in time is deterministic in nature (hence the eventual dependence would be dealt with through deterministic models). Another aspect to underline regarding the above defined processes is that the dependence between univariate processes is only based on them sharing the same multivariate model (e.g. a signal cannot be dependent on another if they \replaced[id = r2]{do not}{don't} share at least one of these models). Although this assumption may restrict the choice of structures of multivariate dependence, this setting is nevertheless sufficiently general to at least well-approximate more complex forms of multivariate latent dependence while providing an important basis to extend this setting in future research. Based on this assumption, we state the following condition:\\
\begin{enumerate}[label=\bfseries (C\arabic*), leftmargin=1cm, leftmargin=1cm]
  \item $\bm{X}_{k, t}$ is independent of $\bm{X}_{k^{\prime}, t}$, $\forall \,\, k \neq k^{\prime}$. \label{cond.indep} \\[0.01cm]
\end{enumerate}
This condition implies that each model that composes a latent model (univariate or multivariate) is independent from the others (e.g. a \ref{WN} process is independent from a \ref{AR} process and two different \ref{AR} processes are independent from each other). The reason for stating this condition is simply to provide proofs on the identifiability of these models for the method that is put forward in this paper (see Sec. \ref{sec:asy_gmwm}). It is of course possible to estimate other forms of dependence in which the processes that compose a latent model are correlated, but this would imply deriving additional theoretical moments and, if not assumed, proving identifiability of these models on a case-by-case basis.

With this premise, in the next sections we present the quantities and relative asymptotic properties that are necessary to define the MGMWM. Using the modelling framework presented in this section, we will then be able to deliver the identifiability of these models through the proposed estimation method and consequently reduce the conditions for its asymptotic properties to some basic regularity conditions.


\section{The Wavelet Cross Covariance}
\label{sec:wccv}

In order to present the proposed MGMWM, we first describe and study the properties of the WCCV. For this purpose, let $J > 0$ be a fixed integer representing the last level of wavelet decomposition of the univariate time series $( X_t^{(i)} )_{t \in \mathbb{Z}}$. This allows us to define the wavelet coefficients for level $j = 1, \dots, J$ as:
\begin{equation*}
  W_{j, t}^{(i)} \equiv \sum_{l = 0}^{L_{j} - 1} h_{j, l} X_{t - l}^{(i)},
\end{equation*}
where $[h_{j, l}]_{l = 0, \dots, L_j-1}$ is the $j^{th}$ level wavelet filter with length $L_j$. In this paper we choose to use the Haar wavelet filter whose length is consequently $L_j = 2^j$. \added[id = r1]{The reason for using this particular filter resides mainly in the fact that, aside from being one of the most commonly used filters in practice, the theoretical form of the WV (see below) for many classes of time series models have been explicitly derived (see e.g. \cite{zhang2008allan}) thereby facilitating the finding of theoretical results presented further on in the paper.}

\replaced[id = r1]{With the above discussion in mind,}{Based on the wavelet coefficients} we can define the WV as:
\begin{equation}
  \eta_{j}^{(i)}(\bm{\theta}^{(i)}) \equiv \text{Var}(W_{j,t}^{(i)}),
\end{equation}
where 
$$\bm{\theta}^{(i)} \equiv \left[\theta_m^{(i)}\right]_{m=1,\hdots,p^{(i)}}$$ is the parameter vector defining the latent model underlying the $i^{th}$ univariate process (of dimension $p^{(i)}$). Indeed, for each time series model there is a corresponding theoretical WV which can be used for estimation purposes in a method-of-moments fashion (as done by the GMWM). In a similar way, we can also define the WCCV \replaced[id = r1]{\cite{whitcher2000wavelet}}{\citep[see][]{whitcher2000wavelet}} at the $j^{th}$ level as follows:
%
\begin{equation}
  \gamma_{j, h}^{(i, i^{\prime})}(\bm{\theta}^{(i,i^{\prime})}) \equiv \text{Cov}\left( W_{j, t}^{(i)}, \, W_{j, t+h}^{(i^{\prime})} \right),
\label{def:theo_crosscov}
\end{equation}
where $h$ represents the lag in time between observations and $\bm{\theta}^{(i,i^{\prime})}$ represents the parameter vector defining the dependence between signals (of dimension $p^{(i,i^{\prime})}$). \added[id = r1]{Given our definitions, the parameter vector of the overall model is therefore $\bm{\theta} \in \bm{\Theta} \subset \mathbb{R}^p$ which can be defined as}
\begin{equation*}
\added[id = r1]{\bm{\theta} \equiv \left[ \left[\bm{\theta}^{(1)}\right]^{\intercal}, \; \ldots ,\; \left[\bm{\theta}^{(I-1,I)}\right]^{\intercal}\right]^{\intercal}},
\end{equation*}
\added[id = r1]{where $p \equiv \sum_{i = 1}^{I} \sum_{j = i}^I p^{(i,j)}$ and where $p^{(i,i)} \equiv p^{(i)}$.}

Having defined these theoretical quantities, we can now define their empirical counterparts that are estimated on the observed signals $(X_t^{(i)})_{t = 1, \dots, T}$ and $({X_t^{(i^{\prime})}})_{t = 1, \dots, T}$, where $T \in \mathbb{N}$. While the unbiased estimator of WV is defined and studied in \cite{percival1995estimation} and \cite{serroukh2000statistical}, here we define the unbiased estimator of WCCV \replaced[id = r1]{\cite{whitcher2000wavelet}}{\citep[see][]{whitcher2000wavelet}} for $h=0$ as: 
%
\begin{equation}
  \hat{\gamma}_{j,0}^{(i, i^{\prime})} \equiv \frac{1}{M_j}\sum_{t = L_j}^{T} W_{j, t}^{(i)} W_{j, t}^{(i^{\prime})},
\label{def:samp_crosscov}
\end{equation}
where $M_j \equiv T-L_j+1$ is the number of wavelet coefficients generated at level $j$. In this paper we limit ourselves to studying the WCCV at lag $h=0$ since this lag is sufficient to identify and estimate the models considered for this work.

Having defined the theoretical and empirical WCCV, we will now study the asymptotic properties of $\hat{\gamma}_{j,0}^{(i, i^{\prime})}$ since they are essential to deliver inference when using the MGMWM which is presented in Sec. \ref{sec:mgmwm}. In \citep[][]{whitcher1999mathematical} the asymptotic normality of $\hat{\gamma}_{j,{0}}^{(i,i^{\prime})}$ was stated for any pair $(i, i^{\prime})$ and any level $j$. However, the conditions for the latter results to hold assume that the process 
$$\left\{\left(W_{j,t}^{(i)}, W_{j,t}^{(i^{\prime})}\right)\right\}_{t \in \mathbb{Z}}$$
is a bivariate Gaussian stationary process which may be a strong assumption to satisfy in general. In this paper we aim at reducing (or weakening) these conditions as well as clearly stating the multivariate asymptotic properties of the WCCV vector. To define the latter, we first simplify notation as follows: $\hat{\gamma}_{j}^{(i,i^{\prime})} \equiv \hat{\gamma}_{j,0}^{(i, i^{\prime})}$ and $\gamma_{j}^{(i,i^{\prime})} \equiv \gamma_{j,0}^{(i, i^{\prime})}(\bm{\theta}^{(i,i^{\prime})})$. Based on this notation, the WCCV vector is defined as:
\begin{equation*}
  \bm{\nu}(\bm{\theta}_0) \equiv \left[ \gamma_j^{(i,i^{\prime})} \right]_{\substack{j = 1,...,J \\ 1 \leq i \leq i^{\prime} \leq I}},
\end{equation*}
with the corresponding vector of estimated WCCV defined as:
\begin{equation*}
  \hat{\bm{\nu}} \equiv \left[ \hat{\gamma}_j^{(i, i^{\prime})} \right]_{\substack{j = 1,...,J \\ 1 \leq i \leq i^{\prime} \leq I}}.
\end{equation*}
In addition, let us also define the first order difference of the $i^{th}$ signal as $(\Delta_t^{(i)})_{t \in \mathbb{Z}}$ (i.e. $\Delta_t^{(i)} \equiv X_t^{(i)} - X_{t-1}^{(i)}$). The reason for considering this process lies in the fact that the wavelet coefficients can be represented as particular linear combinations of the process $(\Delta_t^{(i)})$, i.e.
\begin{equation}
\label{linear}
    W_{j, t}^{(i)} = \bm{c}_j^{\intercal} \bm{\Delta}_{j,t}^{(i)},
\end{equation}
where $\bm{\Delta}_{j,t}^{(i)} \equiv \left(\Delta_{t-L_j+2}^{(i)}, \, \Delta_{t-L_j+3}^{(i)}, \, \dots, \, \Delta_t^{(i)}\right)^{\intercal}$ and $\bm{c}_j$ represents a specific filter vector.
In order to define the first condition, we also define $$\bm{G}(\cdot) = \left(g^{(1)}(\cdot), \, g^{(2)}(\cdot), \, \dots, \, g^{(I)}(\cdot)\right)^{\intercal},$$ as being an $\mathbb{R}^I$-valued measurable function as well as the filtration $\mathcal{F}_t = \sigma(\dots, \, \epsilon_{t-1}, \, \epsilon_t)$ where $\epsilon_t$ are i.i.d. random variables. We can now state the first condition needed for the asymptotic properties of the WCCV vector. \\[0.01cm]
\begin{enumerate}[label=\bfseries (C\arabic*), leftmargin=1cm, resume*]
    \item We assume the multivariate process $(\bm{\Delta}_t)_{t \in \mathbb{Z}}$ is well defined and can be represented as 
    %
    \begin{equation*}
    \label{functional_wold}
        \bm{\Delta}_t \equiv \left(\Delta_{t}^{(1)}, \, \Delta_{t}^{(2)}, \, \dots, \, \Delta_{t}^{(I)}\right)^{\intercal} = \bm{G}(\mathcal{F}_t) .
    \end{equation*}
    \label{stationary}
\end{enumerate}
This condition basically implies that the multivariate process $(\bm{\Delta}_t)$ is ergodic and strictly stationary. Hence, this condition is quite a strong one but is commonly assumed (and often implied by other conditions) to study asymptotic properties in the context of dependent processes. Moreover, this condition is respected by the processes defined in Sec. \ref{sec:multproc} and allows for a very general class of time series models \added[id = r1]{e.g. \cite{zhang2017gaussian}}. 

In order to state the final conditions, we need to deliver additional definitions, starting from $\Delta_t^{\star \, (i) } = g^{(i)}(\mathcal{F}_t^{\star})$ where $\mathcal{F}_t^{\star} = \sigma(\dots, \, \epsilon_0^{\star}, \, \dots, \, \epsilon_{t-1}, \, \epsilon_t)$, where also $\epsilon_0^{\star}$ is an i.i.d. random variable.  Based on this definition, we can notice that for $t < 0$ we have that $\bm{\Delta}_t^{\star} = \bm{\Delta}_t$. Finally, we define $\|Z\|_p \equiv \left( \mathbb{E}|Z|^p\right)^{\nicefrac{1}{p}}$, for $p > 0$, to deliver the other required conditions.\\[0.01cm]
\begin{enumerate}[label=\bfseries (C\arabic*), leftmargin=1cm, resume*]
    \item $\underset{i}{\max} \left\Vert \Delta_{t}^{(i)} \right\Vert_4  < \infty$. \label{cond.mom_bound} \\[0.01cm]
    \item $\underset{i}{\max} \sum_{t=0}^{\infty}\left\Vert \Delta_t^{(i)} - \Delta_t^{\star \, (i) } \right\Vert_4 < \infty$. \label{cond.stability} \\[0.01cm]
\end{enumerate}
Both conditions require the fourth moment of the difference process $(\Delta_t^{(i)})_{t\in\mathbb{Z}}$ (or some function of it) to be finite (for all $i$). More specifically, while Condition \ref{cond.mom_bound} is straightforward to interpret in the latter sense,  Condition \ref{cond.stability} ensures that the cumulative impact of $\epsilon_0^{\star}$ on future values of the process $(\Delta_t^{(i)})_{t\in\mathbb{Z}}$ is finite which allows us to interpret it as a short-range dependence condition \replaced[id = r1]{\cite{wu2011asymptotic}}{\citep[see][]{wu2011asymptotic}}.
We can now deliver the main result of this section and, for this purpose, we further define
$$\bm{W}_t \equiv \left[ W_{j,t}^{(i)}W_{j,t}^{(i^{\prime})} \right]_{\substack{j = 1,...,J \\ 1 \leq i \leq i^{\prime} \leq I}}$$
and the projection operator
    \begin{equation*}
        \mathcal{P}_t(\cdot) \equiv \mathbb{E}(\cdot|\mathcal{F}_t) - \mathbb{E}(\cdot|\mathcal{F}_{t-1}),
    \end{equation*}
which measures how much the conditional expectation of a random variable changes when removing the information from the immediately previous time. Intuitively, this measure should not be large for a ``stable'' signal since these conditional expectations (that contribute to $\mathcal{P}_t(\cdot)$) should be arbitrarily close to the unconditional expectation. Using these last definitions, \added[id = r2]{with ``$\overset{\mathcal{D}}{\to}$'' denoting convergence in distribution,} we can finally state the asymptotic normality of the vector $\hat{\bm{\nu}}$. 
\begin{Theorem}
    \label{thm:asy.nuhat}
    Under Conditions \ref{stationary} to \ref{cond.stability}, and assuming $I,J \in \mathbb{N}$, we have that
    \begin{equation*}
    \sqrt{T}\left(\hat{\bm{\nu}}-\bm{\nu}(\bm{\theta}_0)\right) \replaced[id = r2]{\overset{\mathcal{D}}{\to}}{\xrightarrow[T\rightarrow\infty]{\mathcal{D}}} \mathcal{N}\left(\bm{0}, \bm{V}\right),
    \end{equation*} 
    where $\bm{V} = \mathbb{E}(\bm{D}_0 \bm{D}_0^{\intercal})$ and $\bm{D}_0 \equiv \sum_{t=0}^{\infty} \mathcal{P}_0 \left(\bm{W}_t \right)$.
\end{Theorem}
\vspace{0.3cm}
\begin{proof}
We can split the quantity $\sqrt{T}\left(\hat{\bm{\nu}}-\bm{\nu}(\bm{\theta}_0)\right)$ as follows: $$\sqrt{T}(\tilde{\bm{\nu}} - \bm{\nu}(\bm{\theta}_0)) + \sqrt{T} (\hat{\bm{\nu}} - \tilde{\bm{\nu}}),$$
where $\tilde{\bm{\nu}}$ is an alternative estimator of the WCCV whose elements are given by
\begin{equation*}
    \tilde{\gamma}_j^{(i,i^{\prime})} \equiv \frac{1}{T}\sum_{t=1}^{T} W_{j, t}^{(i)} W_{j, t}^{(i^{\prime})},
\end{equation*}
for all $1 \leq i \leq i^{\prime} \leq I$ and $j = 1,..., J$. Using the latter estimator and Conditions \ref{stationary} to \ref{cond.stability}, we can directly apply the results of \replaced[id = r1]{\cite[Theorem 7]{wu2011asymptotic}}{Theorem 7 in \cite{wu2011asymptotic}} to show asymptotic normality of $\sqrt{T}(\tilde{\bm{\nu}} - \bm{\nu}(\bm{\theta}_0))$ with covariance matrix $ \mathbb{E}(\bm{D}_0 \bm{D}_0^{\intercal})$. Moreover, with $I$ and $J$ being fixed, we have that $\sqrt{T} (\hat{\bm{\nu}} - \tilde{\bm{\nu}}) = \mathcal{O}_p(\nicefrac{1}{T}) = o_p(1)$ and therefore, using Slutsky's theorem, we have that asymptotically $\sqrt{T}\left(\hat{\bm{\nu}}-\bm{\nu}(\bm{\theta}_0)\right)$ is equivalent in distribution to $\sqrt{\mathrm{T}}(\tilde{\bm{\nu}} - \bm{\nu}(\bm{\theta}_0))$ thus concluding the proof. A more detailed explanation of this proof can be found in App. \ref{proof:thm:asy.nuhat}.    
\end{proof}
\vspace{0.1cm}

Having studied the asymptotic properties of the WCCV, the next section develops the MGMWM which makes use of this quantity and its properties for inference.

\section{The Multivariate GMWM}
\label{sec:mgmwm}

Given the results presented in the previous section, we can now introduce the proposed approach to multivariate latent signal modelling which develops from the GMWM \replaced[id = r1]{\cite{guerrier2013wavelet}}{\citep[see][]{guerrier2013wavelet}}. Indeed the latter approach considers a univariate process $(X_t^{(i)})_{t \in \mathbb{Z}}$ and, assuming that this process is generated from a parametric model \replaced[id=r1]{$\bm{F}_{\bm{\theta}^{(i)}}$}{$\bm{F}_{\bm{\theta}}$} known up to the parameter vector \replaced[id=r1]{$\bm{\theta}^{(i)}$}{$\bm{\theta}$}, the GMWM aims at estimating this parameter vector through the following minimization problem:
\iffoo
\else
{\color{black}
\begin{equation}
\label{eq.gmwm}
    \hat{\bm{\theta}}^{(i)} = \argmin_{\bm{\theta} \in \bm{\Theta}^{(i)}} \left(\hat{\bm{\eta}}^{(i)} - \bm{\eta}^{(i)}(\bm{\theta})\right)^{\intercal} \bm{\Omega}^{(i)} \left(\hat{\bm{\eta}}^{(i)} - \bm{\eta}^{(i)}(\bm{\theta})\right),    
\end{equation}
}\fi
where \replaced[id = r1]{$\hat{\bm{\eta}}^{(i)}$ and $\bm{\eta}^{(i)}(\bm{\theta})$}{$\hat{\bm{\eta}}$ and $\bm{\eta}(\bm{\theta})$} are the vectors of estimated and theoretical WV \added[id = r1]{(for the $i$-th process)} respectively, \replaced[id = r1]{$\bm{\Theta}^{(i)} \subset \mathbb{R}^{p^{(i)}}$}{$\bm{\Theta} \subset \mathbb{R}^p$} is the parameter space for \replaced[id=r1]{$\bm{\theta}^{(i)}$}{$\bm{\theta}$} and \replaced[id=r1]{$\bm{\Omega}^{(i)} \in \mathbb{R}^{J \times J}$}{$\bm{\Omega} \in \mathbb{R}^{p \times p}$} is a symmetric positive definite weighting matrix chosen in a suitable manner. For the latter matrix, one choice can be the inverse of the covariance matrix of \replaced[id = r1]{$\hat{\bm{\eta}}^{(i)}$}{$\hat{\bm{\eta}}$}
\replaced[id = r1]{\cite{guerrier2013wavelet}}{\citep[see][]{guerrier2013wavelet}}.  
The consistency and asymptotic normality of \replaced[id = r1]{$\hat{\bm{\theta}}^{(i)}$}{$\hat{\bm{\theta}}$} were shown in \replaced[id = r1]{\cite{guerrier2013wavelet}}{\citet{guerrier2013wavelet}} for any suitable choice of \replaced[id = r1]{$\bm{\Omega}^{(i)}$}{$\bm{\Omega}$}, thereby providing not only a statistically appropriate method for univariate latent time series modelling but also a computationally feasible and efficient approach. 

The idea of the GMWM, being based on the Generalized Method of Moments (GMM) framework, is therefore to inverse the mapping from the WV vector back to the parameter vector $\bm{\theta}$. With this in mind, this idea can be extended by using the vector of WCCV which includes moments that contain information on the subset of parameters in $\bm{\theta}$ that contain information on the latent dependence between signals $i$ and $i^{\prime}$. For this reason, it is possible to build the MGMWM as an extension of the GMWM as follows:
\begin{equation}
\label{eq.mgmwm}
    \hat{\bm{\theta}} = \argmin_{\bm{\theta} \in \bm{\Theta}} \left(\hat{\bm{\nu}} - \bm{\nu}(\bm{\theta})\right)^{\intercal} \bm{\Omega} \left(\hat{\bm{\nu}} - \bm{\nu}(\bm{\theta})\right) .
\end{equation}
Hence, all that is needed to obtain the MGMWM is to postulate a multivariate latent model and compute the empirical WCCV for which we have studied the asymptotic properties in Sec. \ref{sec:wccv}. Moreover, the objective function defined in (\ref{eq.mgmwm}) can be used to test whether considering dependence between sensor errors better fits the observed signals by, for example, comparing it to the distribution of the objective function defined in (\ref{eq.gmwm}) (which could be obtained via parametric bootstrap). However, substituting the WV in the GMWM with the WCCV in the MGMWM is not enough to obtain the necessary inference on the parameter vector $\bm{\theta}$. Indeed, we firstly would need to have an idea on which models it is possible to estimate using this approach and, secondly, use the latter information to then understand the asymptotic properties of this estimator.

Given the above, the next paragraphs will firstly study the so-called \textit{identifiability} of a wide class of multivariate latent time series models and then define the asymptotic properties of the MGMWM.

\subsection{Identifiability of Multivariate Latent Models}
\label{sec.ident}

The condition of model identifiability is essential for any method that aims at estimating its corresponding parameters and is consequently necessary to then obtain its asymptotic properties for inference. In brief, model identifiability through the MGMWM can be defined as
\begin{equation*}
  \bm{\nu}(\bm{\theta}_0) = \bm{\nu}(\bm{\theta}_1) \,\,\; \text{if and only if} \,\,\; \bm{\theta}_0 = \bm{\theta}_1 ,
\end{equation*}
where $\bm{\theta}_0, \bm{\theta}_1 \in \bm{\Theta} \subset \mathbb{R}^p$. This basically means that there is a one-to-one correspondence between the WCCV and the parameter vector $\bm{\theta}$. This property, although essential, is often assumed in practice since it can be considerably challenging to prove.

In this paper however we present results on the identifiability of a class of multivariate latent processes defined in (\ref{MLP}) which can also be used to assess model identifiability on a case-by-case basis. To do so, we need Condition \ref{cond.indep}, which allows us to define the WCCV as the sum of the individual WCCV for each underlying model. In this manner, as mentioned earlier, it is possible to define a general class of multivariate latent models for which we can study the condition of identifiability. It must be noted that, in this paper, we use the word ``class'' to define an overall model which includes all possible subsets of model combinations that compose this overall model (or class). Given this definition, we now study the identifiability of the following general class of multivariate latent models:\\[0.01cm]
\begin{enumerate}[label=\bfseries (M\arabic*), leftmargin=1.2cm]
    \item $\bm{X}_t \equiv \bm{X}_{1,t} + \bm{X}_{2,t} + \bm{X}_{3,t} + \bm{X}_{4,t}$. \label{mod1} \\[0.01cm]
    \item $\bm{X}_t \equiv \bm{X}_{1,t} + \bm{X}_{3,t} + \sum_{k = 5}^{K} \bm{X}_{k,t}$. \label{mod2} \\[0.01cm]
\end{enumerate}
In the univariate equivalent setting, these two classes of models are commonly used in various disciplines including finance, economics and engineering \replaced[id = r1]{\cite{el2008analysis,granger1976time,skrondal2007latent}}{\cite[see][]{el2008analysis,granger1976time,skrondal2007latent}}. More specifically, the class of multivariate latent models \ref{mod1} consists of processes that are not second order stationary, although their first-order backward difference is indeed so, while the \ref{mod2} class is stationary based on the parameter definitions of the models given earlier. Considering these general classes of \replaced[id = r2]{multivariate}{mutlivariate} latent models (which therefore include all possible sub-model combinations), the following proposition states injectivity of $\bm{\nu}(\bm{\theta})$ for the first class of models.

\begin{Proposition}
\label{prop:mod1.injective}
    Under Condition \ref{cond.indep}, we have that $\bm{\nu}(\bm{\theta})$ is an injective function of $\bm{\theta}$ for the class of models \ref{mod1}.
\end{Proposition}
\vspace{0.3cm}
\begin{proof}
   The proof is based on verifying the conditions for identifiability in \replaced[id = r1]{\cite[Theorem 2]{komunjer2012global}}{Theorem 2 of \cite{komunjer2012global}}. To do so, we start by defining $\bm{\nu}^{\star}(\bm{\theta})$ which is an appropriately chosen $p$-dimensional subvector of $\bm{\nu}(\bm{\theta})$ (in such a way that both $\bm{\nu}^{\star}(\bm{\theta})$ and $\bm{\theta}$ belong to $\mathbb{R}^p$). If the subvector $\bm{\nu}^{\star}(\bm{\theta})$ is identifiable, we have that also the entire vector $\bm{\nu}(\bm{\theta})$ is identifiable. Using log-transformations of the parameters (which is a one-to-one function), we can easily prove Assumptions A and C required by \cite{komunjer2012global} which pertain to the behaviour of the function $\bm{\nu}^{\star}(\bm{\theta})$ (i.e. twice-differentiable and divergent when $\bm{\theta}$ diverges). Finally, we have that the determinant of the Jacobian of $\bm{\nu}^{\star}(\bm{\theta})$ is non-zero which verifies Assumptions B and D in \cite{komunjer2012global} and proves the identifiability of $\bm{\nu}(\bm{\theta})$. A more detailed explanation of this proof can be found in App. \ref{proof:prop:mod1.injective}.
\end{proof}
\vspace{0.3cm}

This is an important result since it provides the fundamental basis for this class of models to be identifiable for the MGMWM. However, directly showing the injectivity of $\bm{\nu}(\bm{\theta})$ for the class of models \ref{mod2} is not obvious since explicitly verifying whether the determinant of the Jacobian of $\bm{\nu}(\bm{\theta})$ is non-zero is difficult. Nevertheless, we know that the WCCV has a direct relationship with the cross-covariance function through the cross-spectral density and we can therefore understand at least if the cross-covariance function is a one-to-one function thereby providing insight to the identifiability of $\bm{\nu}(\bm{\theta})$. For this purpose, let us denote the cross-covariance function as
$$\bm{C}(\bm{\theta}) \equiv \left[ C_l^{(i, i^{\prime})}(\bm{\theta}) \right]_{\substack{l \in \mathbb{Z} \\ 1 \leq i \leq i^{\prime} \leq I}},$$
where $C_l^{(i, i^{\prime})}(\bm{\theta}) \equiv \cov\left( X_{t}^{(i)}, \, X_{t+l}^{(i^{\prime})} \right)$ is the cross-covariance with respect to the directed lag $l$.
Notice that $C_l^{(i, i^{\prime})}(\bm{\theta}) \neq C_{-l}^{(i, i^{\prime})}(\bm{\theta})$ in general.

Considering these quantities, we firstly show the injectivity of $\bm{C}(\bm{\theta})$ for the class of models \ref{mod2} and, taking advantage of the one-to-one correspondence between $\bm{C}(\bm{\theta})$ and \added[id = r1]{the cross-spectral density, denoted as} $\bm{S}(\bm{\theta})$, study the one-to-one correspondence between $\bm{S}(\bm{\theta})$ and $\bm{\nu}(\bm{\theta})$. In order to do so, we first need to define the following condition where we let $\phi_k^{(i)}$ denote the parameter of the $k^{th}$ \ref{AR} process in the $i^{th}$ signal. \\[0.01cm]
\begin{enumerate}[label=\bfseries (C\arabic*), leftmargin=1cm, resume]
  \setcounter{enumi}{4}
    \item If the $i^{th}$ time series consists in the sum of $(K-4)$ \ref{AR} processes, for $K \geq 5$, then we have that all $\phi_k^{(i)}$ are distinct and nonzero.
    \label{mod:distinct} \\[0.01cm]
\end{enumerate}
This condition ensures that the values of the autoregressive parameters of \ref{AR} processes contributing to the $i^{th}$ signal are distinct since otherwise the theoretical cross-covariance function of this signal is not injective. Having given this condition, we can now deliver the following lemma which consists in the first step to proving injectivity of $\bm{\nu}(\bm{\theta})$.

\begin{Lemma}
    \label{lem:ccv.injec}
    Under conditions \ref{cond.indep} and \ref{mod:distinct}, we have that the cross-covariance function $\bm{C}(\bm{\theta})$ of \ref{mod2} is injective.
\end{Lemma}
\vspace{0.3cm}
\begin{proof}
The proof of this lemma is the same as that for Proposition \ref{prop:mod1.injective} except that the steps are made for \deleted[id = r2]{the} an appropriate sub-vector $C^{\star}(\bm{\theta})$ of the autocovariance sequence $C(\bm{\theta})$. For more details see App. \ref{proofs.ident}.
\end{proof}
\vspace{0.3cm}

This result is already an important finding since the cross-covariance function is used as a means to estimate these processes via estimators such as the MLE or the GMM. Hence, the identifiability of these models can already easily be proven based on Lemma \ref{lem:ccv.injec}. Nevertheless, as underlined earlier, proving the same property directly for $\bm{\nu}(\bm{\theta})$ is extremely complicated and we will therefore assume an additional condition for this reason. In this sense, let us express $\bm{\nu}(\bm{\theta}) \equiv \bm{f}(\bm{S}(\bm{\theta}))$ where $\bm{f}(\cdot)$ is a known vector function \added[id = r1]{e.g. \cite{el2008analysis}}.\\[0.01cm]
\begin{enumerate}[label=\bfseries (C\arabic*), leftmargin=1cm, resume]
  \setcounter{enumi}{5}
    \item $\bm{f}(\cdot)$ is an injective function.
    \label{cond.s_to_wv}\\[0.01cm]
\end{enumerate}
Regarding the above condition, existing results for univariate time series \added[id = r1]{e.g. \cite{greenhall1998spectral, guerrier2016identifiability}} underline how this condition is not necessarily a strong one to assume in general. \added[id = r1]{Indeed, \cite{guerrier2016identifiability} directly show the identifiability of many latent models including ARMA models and discuss general identifiability for other models supported by the findings of \cite{greenhall1998spectral} who, in the continuous-time case, shows that this condition is almost always verified when using the Haar WV.} With the addition of the latter condition, we can now deliver the following proposition.
\begin{Proposition}
    \label{prop:mod2.injective}
    Under conditions \ref{cond.indep}, \ref{mod:distinct} and \ref{cond.s_to_wv}, $\bm{\nu}(\bm{\theta})$ is an injective function of $\bm{\theta}$ for the class of models \ref{mod2}.
\end{Proposition}
\vspace{0.3cm}
\begin{proof}
The proof of this proposition is a straightforward consequence of Lemma \ref{lem:ccv.injec} since the cross-spectral density function is an injective function of the cross-covariance function $C(\bm{\theta})$ and, using Condition \ref{cond.s_to_wv}, we prove Proposition \ref{prop:mod2.injective}.   
\end{proof}
\vspace{0.3cm}

Based on Propositions \ref{prop:mod1.injective} and Proposition \ref{prop:mod2.injective}, we have provided the results necessary to determine the asymptotic properties of the proposed MGMWM estimator to hold for two very general classes of models (based on the defined conditions) and identifiability can be reasonably assumed for other more specific classes of models if Condition \ref{cond.s_to_wv} holds.

\subsection{Asymptotic Properties of the MGMWM}
\label{sec:asy_gmwm}

Having studied the properties of the WCCV in Sec. \ref{sec:wccv} and discussed the identifiability of different classes of multivariate latent models in Sec. \ref{sec.ident}, we can now study the asymptotic properties of the MGMWM. To do so, let us define 
\begin{equation*}
 Q_{T}( \bm{\theta} ) \equiv ||\hat{\bm{\nu}} - \bm{\nu} (\bm{\theta})||_{\bm{\Omega}_T} =  (\hat{\bm{\nu}} - \bm{\nu} (\bm{\theta}))^{\intercal} \bm{\Omega}_T (\hat{\bm{\nu}} - \bm{\nu} (\bm{\theta}))
\end{equation*}
as the objective function of the MGMWM in (\ref{eq.mgmwm}) where $\bm{\Omega}_T$ is an estimated positive definite matrix. In a similar way, we can define its theoretical counterpart as 
$$Q_0(\bm{\theta}) \equiv ||\bm{\nu}(\bm{\theta}_0) - \bm{\nu} (\bm{\theta})||_{\bm{\Omega}},$$
with $\bm{\theta}_0$ and $\bm{\Omega}$ being the true parameter vector and a fixed positive definite weighting matrix respectively.

Considering these definitions, let us further define $||\cdot||_S$ as being the matrix spectral norm \added[id = r2]{and ``$\overset{p}{\to}$'' denoting convergence in probability} which allow us to state a set of conditions that are necessary to obtain desirable asymptotic properties for the MGMWM.\\[0.01cm]
\begin{enumerate}[label=\bfseries (C\arabic*), leftmargin=1cm, resume]
  \setcounter{enumi}{6}
    \item $\bm{\Theta}$ is compact. \label{compact} \\[0.01cm]
    \item $|| \bm{\Omega}_T - \bm{\Omega}||_S \overset{p}{\to} 0$. \label{mat.elem} \\[0.01cm]
    \item $\bm{\nu}(\bm{\theta}) = \bm{\nu}(\bm{\theta}_0)$, if and only if $\bm{\theta} = \bm{\theta}_0$. \label{ident} \\[0.01cm]
    \item $Q_0(\bm{\theta})$ is continuous $\forall \, \bm{\theta} \in \bm{\Theta}$. \label{obj.cont} \\[0.01cm]    
\end{enumerate}
Conditions \ref{compact}, \ref{mat.elem} and \ref{obj.cont} are standard regularity conditions that, among others, ensure the uniform convergence of $Q_T(\bm{\theta})$ while Condition \ref{ident} is related to identifiability which was discussed for different classes of models in the previous section (hence this condition was proven for the general classes of Models \ref{mod1} and \ref{mod2}). Based on these conditions we can state the following theorem.\vspace{0.3cm}
\begin{Theorem}
\label{thm:gmwm.consis}
  Under conditions \ref{stationary}-\ref{cond.stability} and \ref{compact}-\ref{obj.cont}, we have that 
  \begin{equation*}
    \hat{\bm{\theta}} \overset{\replaced[id = r2]{p}{P}}{\to} \bm{\theta}_0 .
  \end{equation*}
\end{Theorem}
\vspace{0.3cm}
\begin{proof}
By replacing the terms in the difference $Q_T(\bm{\theta}) - Q_0(\bm{\theta})$ by their expressions and rearranging terms we can bound $\underset{\bm{\theta}}{\sup}|Q_T(\bm{\theta}) - Q_0(\bm{\theta})|$ by a series of terms that depend on the parameter $\bm{\theta}$ and on the difference between weighting matrices $\bm{\Omega}_T$ and $\bm{\Omega}$ which, based on Theorem \ref{thm:asy.nuhat} as well as Conditions \ref{compact} and \ref{mat.elem}, represent bounded elements multiplied by a term that goes to zero. Hence, the uniform convergence of $Q_T(\bm{\theta})$ is verified and, using \replaced[id = r1]{\cite[Theorem 2.1]{newey1994large}}{Theorem 2.1 in \cite{newey1994large}}, we combine this result with Conditions \ref{ident} and \ref{obj.cont} thereby proving Theorem \ref{thm:gmwm.consis}. A more detailed proof of this theorem can be found in App. \ref{proof:thm:gmwm.consis}
\end{proof}
\vspace{0.3cm}

Having proven consistency of the MGMWM, we now discuss its asymptotic normality. To do so, let us define
$$\bm{A}(\bm{\theta}_0) \equiv \frac{\partial}{\partial \bm{\theta}^{\intercal}} \bm{\nu}(\bm{\theta})\Big|_{\bm{\theta} = \bm{\theta}_0}$$
and consider the following additional condition:\\[0.01cm]
\begin{enumerate}[label=\bfseries (C\arabic*), leftmargin=1cm, resume]
    \item $\bm{\theta}_0 \mbox{ is an interior point of }\bm{\Theta}$, where $\bm{\Theta} \subset \mathbb{R}^p$ is convex. \label{interior} \\[0.01cm]
    \item $\bm{A}(\bm{\theta}_0)^{\intercal} \bm{\Omega}\bm{A}(\bm{\theta}_0)$ is non-singular. \label{asy.non-singular} \\[0.01cm]
\end{enumerate}
Condition \ref{interior} allows us \replaced[id = r1]{to make use of the multivariate mean value theorem}{to perform a MacLaurin expansion} while Condition \ref{asy.non-singular} is another regularity condition which \replaced[id = r1]{allows an asymptotic covariance matrix of $\hat{\bm{\theta}}$ to be defined}{allows to define an asymptotic covariance matrix of $\hat{\bm{\theta}}$}. Having these conditions, we can deliver the final theoretical result of this paper.
\begin{Theorem}
\label{thm:gmwm.asy}
  Under conditions \ref{stationary}-\ref{cond.stability} and \ref{compact}-\ref{asy.non-singular}, we have that
\begin{equation*}
  \sqrt{T} \, \left( \hat{\bm{\theta}} - \bm{\theta}_0 \right) \replaced[id = r2]{\overset{\mathcal{D}}{\to}}{\xrightarrow[T\rightarrow\infty]{\mathcal{D}}} \mathcal{N}(\bm{0},\bm{\Xi}),
\end{equation*} 
where $\bm{\Xi}$ is the asymptotic covariance matrix.\\
\end{Theorem}
\vspace{0.3cm}
\begin{proof}
Having proven consistency in Theorem \ref{thm:gmwm.consis}, we can use this result to prove asymptotic normality. Indeed, by definition we have that $\nicefrac{\partial}{\partial \bm{\theta}} \, Q_T(\bm{\theta}) |_{\bm{\theta} = \hat{\bm{\theta}}} = \bm{0}$ and, making use of the multivariate mean value theorem, we can re-express and rearrange this last equality to obtain
\begin{align*}
    \sqrt{T}\left( \hat{\bm{\theta}} - \bm{\theta}_0 \right) 
            =& -\left[\bm{B}(\hat{\bm{\theta}})\,\bm{\Omega}_T \,\bm{A}(\hat{\bm{\theta}}, \bm{\theta}_0)\right]^{-1} \\
            & \times \bm{B}(\hat{\bm{\theta}})\,\bm{\Omega}_T \,\sqrt{T} \left( \hat{\bm{\nu}} - \bm{\nu}(\bm{\theta}_0) \right),
\end{align*}
where $\bm{B}(\hat{\bm{\theta}})$ and $\bm{A}(\hat{\bm{\theta}}, \bm{\theta}_0)$ are both matrices that converge in probability to $\bm{A}(\bm{\theta}_0)^{\intercal}$ and $\bm{A}(\bm{\theta}_0)$ respectively  based on Theorem \ref{thm:gmwm.consis}. Using Slutsky's theorem and Theorem \ref{thm:asy.nuhat} we obtain the final result of Theorem \ref{thm:gmwm.asy}. A more detailed proof can be found in App. \ref{proof:thm:gmwm.asy} where the explicit form of the asymptotic covariance matrix $\bm{\Xi}$ is also given.
\end{proof}
\vspace{0.3cm}

With the above results we have now delivered an important framework to estimate and provide inference on a wide range of multivariate latent models that can considerably extend the modelling possibilities for inertial sensor calibration as well as for a variety of different applied settings. \added[id = r1]{In order to make use of these results in the following sections, some computational approaches are needed to obtain the MGMWM estimator in practice. Various approaches can be used for this purpose, however we recommend using Algorithm \ref{algo:mgmwm}.}

\iffoo
\else
{\color{black}
\begin{algorithm}
 1. Compute empirical WCCV $\hat{\bm{\nu}}$ for a suitable number of scales $J < \log_2(T)$ (in practice we recommend using $J = \floor{\log_2(T)} - 1$)\;
 \BlankLine
 {2. Compute starting value} $\tilde{\bm{\theta}}$:
 
 \For{$i\leftarrow 1$ \KwTo $I$}{
    \For{$j\leftarrow i$ \KwTo $I$}{
        \eIf{i == j}{
        Compute $\bm{\Omega}^{(i)}$ using the approach considered in \cite{guerrier2013wavelet}\;
        Compute $\tilde{\bm{\theta}}^{(i)}$ (based on $\bm{\Omega}^{(i)}$) as defined in (\ref{eq.gmwm}) using the algorithm described in \cite{balamuta2017computationally}
        }{
        $\tilde{\bm{\theta}}^{(i,j)} \longleftarrow \mathbf{0}$}
    }
 }
 \BlankLine
 
 {3. Compute} $\hat{\bm{\theta}}$:
 \begin{enumerate}
     \item  Estimate $\bm{\Omega}$ by $\bm{\Omega}_T$ (for example taking the inverse\\ of the covariance matrix estimator \cite{andrews1991heteroskedasticity})$^\ddag$.
     \item  Compute $\hat{\bm{\theta}}$ using $\bm{\Omega}_T$ and the starting value $\tilde{\bm{\theta}}$) as defined in (\ref{eq.mgmwm}) using the algorithm described in \cite{balamuta2017computationally}
 \end{enumerate}
 
 \caption{MGMWM Estimation Procedure. $^\ddag$When the dimension is large, the inverse of a large positive-definite matrix is computationally expansive. In this case, we recommend choosing $\bm{\Omega}$ to be the inverse of the diagonal matrix constructed by the diagonal elements of the covariance matrix of $\hat{\bm{\nu}}$}
 \label{algo:mgmwm}
\end{algorithm}
}
\fi
 \vspace{0.25cm}
 \added[id = r1]{Making use of this approach,} the next sections highlight the good properties of the proposed MGMWM framework through some simulation studies and show how these results can be of great importance for considering dependence between sensors which, to date, has not been simple to model despite its relevance for enhanced navigation precision.

\section{Simulation study}
\label{sec:simulation}
In this section we present a simulation study which replicates the simulation setting considered in \cite{vaccaro2017reduced} who propose an approach to estimate a specific multivariate model composed of a \ref{WN} and \ref{RW} model where multivariate dependence is only present through the latter (a second simulation study is presented in App. \ref{sim2} where a more complex setting is considered and for which the method in \cite{vaccaro2017reduced} cannot be applied). Within this setting we compare our approach with the estimators proposed in \cite{vaccaro2017reduced} where they consider their Generalized Least Square (GLS) approach using different scales of the Allan Variance (AV). In the latter perspective, for their simulations they consider the GLS discarding either the last two scales of AV (GLS(2)) or the last three scales (GLS(3)). As for the type of simulated process, \cite{vaccaro2017reduced} considered a multivariate process for an array of six gyroscopes constituted by, as mentioned above, the sum of \ref{WN} and \ref{RW} processes (i.e. multivariate WN and RW processes). For the sake of clarity and interpretation of results, in this section we only consider an array of three gyroscopes (corresponding to the first three gyroscopes in their simulation setting). For this reason, the covariance matrix $\bm{\Sigma}$ of the multivariate WN process is given by
\begin{equation*}
\bm{\Sigma} = 
  \begin{bmatrix}
      \sigma_1 & 0 & 0\\ 0 & \sigma_2 & 0\\ 0 & 0 & \sigma_3
  \end{bmatrix}=
  10^{-3}\begin{bmatrix}
      0.1010 & 0 & 0\\ 0 & 0.0712 & 0\\ 0 & 0 & 0.0490
  \end{bmatrix},
\end{equation*}
while the covariance matrix $\bm{\Lambda}$ of the multivariate RW process is given by
\begin{equation*}
\bm{\Lambda} = 
  \begin{bmatrix}
      \lambda_{11} & \lambda_{12} & \lambda_{13}\\ \lambda_{12} & \lambda_{22} & \lambda_{23}\\ \lambda_{13} & \lambda_{23} & \lambda_{33}
  \end{bmatrix}=
  \begin{bmatrix}
      \phantom{-}0.0119 & -0.0004 & \phantom{-}0.0048\\ -0.0004 & \phantom{-}0.0220 & \phantom{-}0.0093\\ \phantom{-}0.0048 & \phantom{-}0.0093 & \phantom{-}0.1628
  \end{bmatrix}.
\end{equation*}
The notation used for the parameters of these matrices has been slightly modified for simplicity (i.e. $\sigma_i \equiv \sigma^{(i,i)}$ and $\lambda_{ii^{\prime}} \equiv \lambda^{(i,i^{\prime})}$). Based on these definitions, the univariate WN processes are uncorrelated (hence three parameters to \replaced[id = r2]{characterize}{characterise} their covariance matrix) while the RW processes are all correlated (hence a full symmetric matrix with six parameters to \replaced[id = r2]{characterize}{characterise} it).

\begin{figure}
  \centering
  \includegraphics[scale = 0.55]{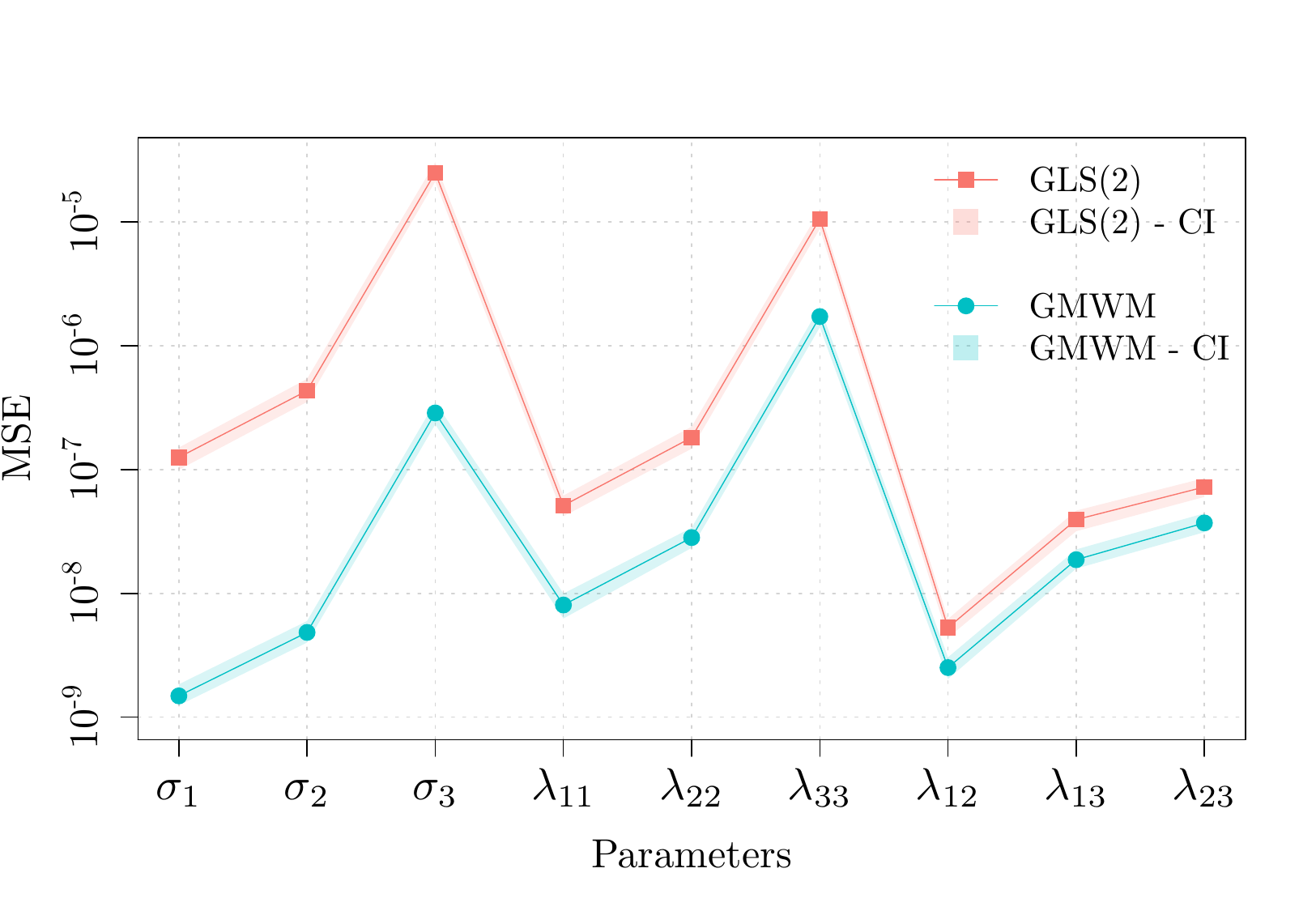}
  \caption{Solid lines represent the sample MSE of GLS(2) and GMWM estimators for a multivariate WN + RW process based on $500$ replications and sample size $T=10^5$. Lower and upper dotted lines represent respectively the bootstrap $2.5\%$ and $97.5\%$ confidence intervals.}
  \label{fig:1}
\end{figure}

In order to study the finite sample performance of the estimators we make use of the empirical Mean Squared Error (MSE) which takes into account bias and variance of an estimator. This measure is based on 500 simulated replicates where the estimators are computed on each multivariate simulated process of length $T = 10^5$. The \replaced[id = r2]{empirical}{empricial} MSE of the estimators are shown in Fig. \ref{fig:1} where, for the GLS estimators, only the results for GLS(2) were represented since they were visually indistinguishable from the results of GLS(3).

Considering that the shaded areas represent the empirical confidence intervals of the MSE, it is evident from the plots that the MGMWM outperforms the GLS(2) (as well as the omitted GLS(3)) for the estimation of all the parameters considered in the simulation study (the corresponding boxplots of the results can be found in App. \ref{sim2}). As mentioned at the beginning of this section, another simulation study was performed on more complex composite processes for which other methods (including the GLS approach) are either unavailable or computationally unfeasible (these simulations can be found in App. \ref{sim2}).

\section{Case study}
\label{sec:applications}

In this section we apply the MGMWM to perform a multisensor calibration procedure on some real data. Sensor calibration is an extremely important procedure to ensure navigation performance, especially for low-cost IMUs whose employment is rapidly increasing but suffers from more complex error measurements that need to be accounted for. In general, this procedure is performed independently on each sensor composing the IMU which implicitly assumes that the error measurements of each sensor are independent of each other. The latter unrealistic assumption has nevertheless been necessary to deal with the complex error signals of these sensors which have been extremely difficult to characterize also at an individual level. Since the GMWM has \replaced[id = r1]{allowed the problems to be greatly reduced}{allowed to greatly reduce the problems} in modelling the individual error measurements of each sensor, the MGMWM is an extension that \replaced[id = r1]{allows complex multivariate dependencies between sensors to be taken into account}{allows to take into account complex multivariate dependencies between sensors} and consequently make the modelling procedure more realistic and improve navigation precision. Moreover, this approach is starting to be taken increasingly into account as underlined in \replaced[id = r1]{\cite{vaccaro2017reduced}}{\citet{vaccaro2017reduced}} where the construction of virtual gyroscopes based on sensor interdependence can considerably reduce errors in different navigation measurements.

For illustrative purposes, in this section we only consider two stochastic error measurements coming from the accelerometer X-axes of an array of two same-model IMUs (namely XSens MTi-G IMUs) sampled at 250Hz over roughly 1 hour with consequent sample size $T = 873'684$ (this procedure can of course take into account more sensors).

\begin{figure*}
  \centering
  \includegraphics[width=0.9\textwidth]{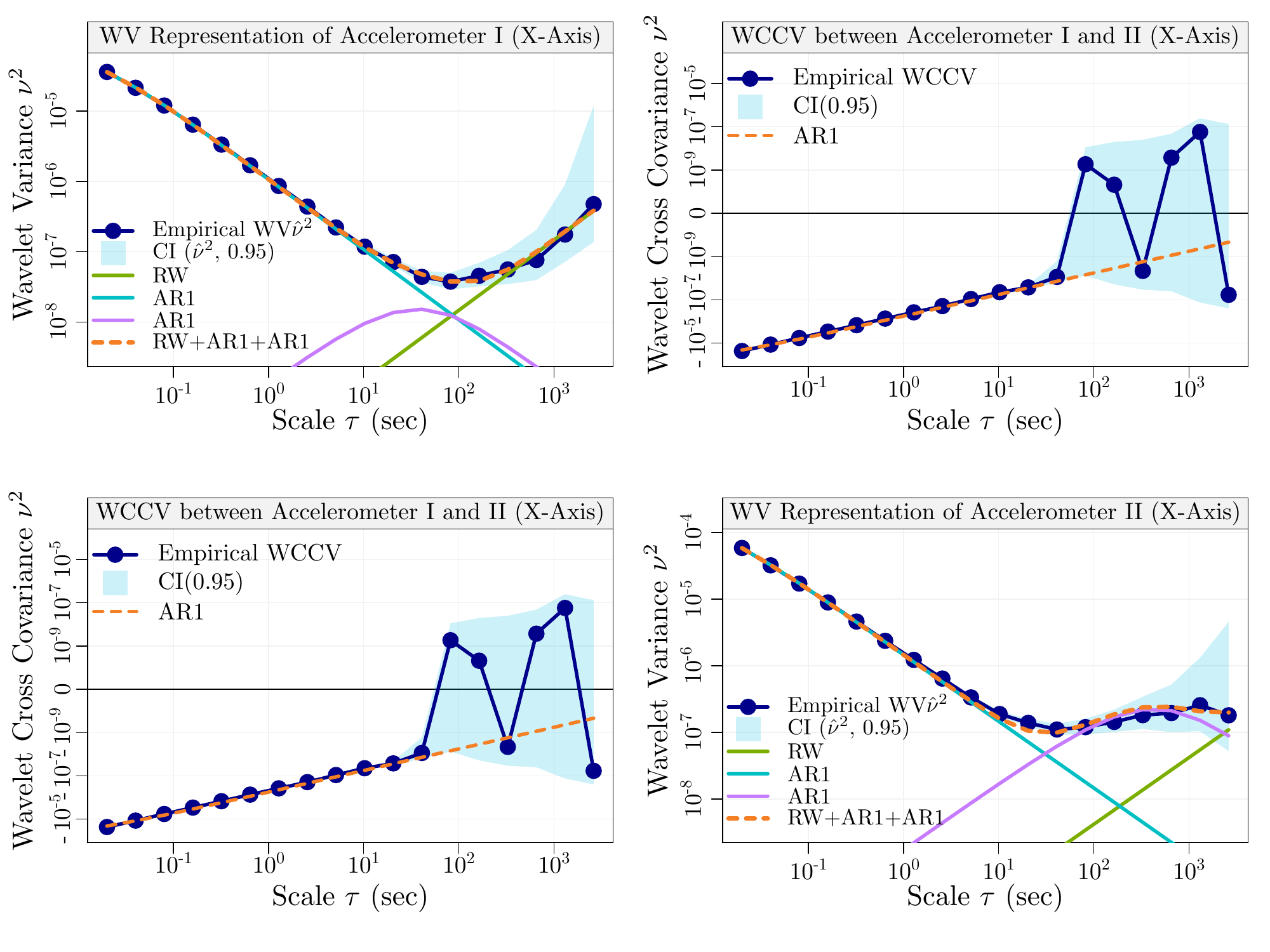}
  \caption{Diagonal Plots: \added[id = r2]{Logarithm of the} estimated WV of accelerometer error signal on the X-Axes of the first IMU (top-left) and the second IMU (bottom-right) with \added[id = r2]{the logarithm of the} WV implied by the MGMWM estimator. Off-Diagonal Plots: \added[id = r2]{Logarithm of the absolute value of the} estimated WCCV \added[id = r2]{ (multiplied by its original sign) and the logarithm of the} WCCV implied by the MGMWM estimator.}
  \label{fig:2}
\end{figure*}

The diagonal plots of Fig. \ref{fig:2} (top-left and bottom-right) show the log-log plots of the empirical WV for each error signal of these sensors where the solid dotted lines represent the estimated WV (surrounded by a shaded area representing the confidence intervals). Simply looking at the lines in the diagonal plots we can see how they appear to have similar behaviour with different patterns at larger scales (as expected) where there is more variability. \replaced[id = r1]{In order to select the appropriate multivariate model we first focus on the individual signals making use of visual matching of the WV plots and selection criteria based on the WV (see \cite{guerrier2015automatic}).}{After testing different models}  \added[id = r1]{Based on this,} we find that the sum of two \ref{AR} models (AR(1)) and a \ref{RW} model (RW) fits both signals well as seen by how close the dashed orange line follows the empirical WV. The contribution of the individual models is represented by the individual solid lines below the empirical WV. In practice, the first AR(1) process would usually be replaced by a \ref{WN} model (WN) since the slope of the latter is close to the one observed at the first scales in the plots and the WN is more straightforward to integrate within a navigation filter. In addition, within the latter filter an AR(1) process with a very small correlation time is unlikely to be observable. However, for the purpose of this study we consider the models that better fit the observed error signals which obviously does not exclude the possibility of considering other models in practice. Having defined the individual models, we can now understand if any of these individual models contributes to the WCCV represented by the solid dotted line in the off-diagonal plots of Fig. \ref{fig:2} (top-right and bottom-left). Estimating all possible dependencies, we found that one of the \ref{AR} models \replaced[id = r1]{allowed the WCCV to be well described}{allowed to well describe the WCCV} as shown by the orange dashed line representing the WCCV implied by the estimated multivariate model.

This case study highlights how the MGMWM \replaced[id = r1]{allows to be increased}{allows to increase} the ease with which complex multivariate composite processes can be modelled, improving in terms of statistical performance and flexibility over existing approaches for multivariate sensor calibration. Indeed, its generality and computational feasibility can greatly contribute, for example, to facilitating and improving the creation of virtual sensors that can considerably enhance navigation performance. Indeed, this approach has already been used in \replaced[id = r1]{\cite{zhang2018optimal}}{\citet{zhang2018optimal}} to construct an optimal virtual gyroscope with good results both in simulated and applied settings.

\section{Conclusions}
\label{sec:conclusions}

This paper has presented a new approach to modelling a wide class of (composite) multivariate time series in order to take into account complex dependence structures between individual signals, thereby providing a contribution to improve multivariate time series modelling in general. Building upon the GMWM framework, we studied the WCCV and extended the asymptotic properties of its estimator to the more general multivariate case under weaker conditions compared to existing work. Based on these properties, we extended the GMWM framework by proposing the MGMWM which takes into account the WCCV to describe the dependence between individual time series. Benefitting from different identifiability results, we delivered the asymptotic properties of the MGMWM estimators under relatively weak conditions and highlighted its improved performance with respect to existing approaches. The case study underlined how the MGMWM can greatly support the calibration of the multivariate stochastic error signals coming from IMUs and consequently contribute to improving navigation performance by, for example, testing for dependence between sensors, modelling inter-axis dependence within an IMU or across sensors in an array of redundant sensors and, based on the latter, constructing optimal virtual sensors.

\bibliographystyle{unsrt}
\FloatBarrier\bibliography{ref} 

\AtEndDocument{

\section*{Supplementary material}

APPENDIX

*A. Proof of Theorem 1*

We prove the asymptotic normality of the vector of WCCV $\hat{\boldsymbol{\nu}}$ in a two-step fashion. Indeed, we intend to study the properties of the quantity $\sqrt{T}\left(\hat{\boldsymbol{\nu}} - \boldsymbol{\nu}(\boldsymbol{\theta}_0)\right)$ which we can re-express as

$$\sqrt{T}\left(\hat{\boldsymbol{\nu}} - \boldsymbol{\nu}(\boldsymbol{\theta}_0)\right) = \sqrt{T}(\tilde{\boldsymbol{\nu}} - \boldsymbol{\nu}(\boldsymbol{\theta}_0)) + \sqrt{T}(\hat{\boldsymbol{\nu}} - \tilde{\boldsymbol{\nu}}), \tag{1}$$

where $\tilde{\boldsymbol{\nu}}$ is an alternative estimator of the WCCV whose elements are given by

$$\tilde{\gamma}_j^{(i,i')} \equiv \frac{1}{T}\sum_{t=1}^{T} W_{j,t}^{(i)} W_{j,t}^{(i')},$$

for all $1 \leq i \leq i' \leq I$ and $j = 1, ..., J$. We can notice that this estimator assumes that the wavelet decomposition delivers $T$ wavelet coefficients at each scale of decomposition $j$ (instead of $T - L_j + 1$) which is not the case in practice. However, this alternative estimator only plays a role in proving the asymptotic properties of $\hat{\boldsymbol{\nu}}$ and is consequently never assumed to be applied in practice. Indeed, the goal of this two-step proof is to show asymptotic normality of the quantity $\sqrt{T}(\tilde{\boldsymbol{\nu}} - \boldsymbol{\nu}(\boldsymbol{\theta}_0))$, on one side, and the convergence to zero of the quantity $\sqrt{T}(\hat{\boldsymbol{\nu}} - \tilde{\boldsymbol{\nu}})$ which will allow us to use Slutsky's theorem to obtain the properties of the original quantity of interest $\sqrt{T}\left(\hat{\boldsymbol{\nu}} - \boldsymbol{\nu}(\boldsymbol{\theta}_0)\right)$ based on the decomposition in (1).

Considering the above discussion and defining

$$\boldsymbol{W}_t \equiv \left[W_{j,t}^{(i)} W_{j,t}^{(i')}\right]_{\substack{j=1,...,J \\ 1 \leq i \leq i' \leq I}},$$

the first step of this proof is given by the following lemma.

LEMMA A.1: *Under Conditions (C2)-(C4) and with both $I \in \mathbb{N}$ and $J \in \mathbb{N}$, we have that*

$$\sqrt{T}\left(\tilde{\boldsymbol{\nu}} - \boldsymbol{\nu}(\boldsymbol{\theta}_0)\right) \xrightarrow{\mathcal{D}} \mathcal{N}\left(\boldsymbol{0}, \mathbb{E}(\boldsymbol{D}_0\boldsymbol{D}_0^T)\right),$$

*where $\boldsymbol{D}_0 \equiv \sum_{t=0}^{\infty}\mathcal{P}_0\left(\boldsymbol{W}_t\right)$.*

*Proof.* The proof of asymptotic normality of $\hat{\boldsymbol{\nu}}$ is based on the requirements of [1, Theorem 7] which provides a central limit theorem result for the sample autocovariance function. The latter uses an alternative representation of causal stationary ergodic processes in order to deliver weaker (and more verifiable) conditions to prove the weak convergence theorem in [2] (see [1,

Theorem 3]). Based on these results, the asymptotic normality of the sample (auto)covariance vector is proven through some basic steps which require the verification that certain quantities are bounded. These steps are the same also for the sample WCCV $\bar{\nu}$ which is a covariance function between jointly stationary time series $(W_{j,t}^{(i)})$ and $(W_{j,t}^{(i')})$ (based on Condition **(C2)**) thereby respecting the requirements of [1, Theorem 7]. In the latter, they focus on the quantity

$$T_n = \sum_{i=1}^{n} X_i X_{i+j} - n\gamma_j,$$

where $n$ represents the sample size ($T$ in our case) and $\gamma_j$ represents the true autocovariance at lag $j$ (not to be confused in terms of notation with the wavelet decomposition $j$ in the present paper). In our case, the lag $j$ is replaced by $i'$ since we don't consider lagged observations but only concurrent observations from two different (but jointly stationary) time series $(W_{j,t}^{(i)})$ and $(W_{j,t}^{(i')})$. To adapt the notation, in our case let us define the set $\mathcal{I} \equiv \{(i, i', j) \in I^2 \times J \mid i \leq i'\}$ as being the set of distinct combinations of time series and levels of wavelet decomposition. Hence, $\iota \in \mathcal{I}$ is an element of this set which therefore represents one of these possible combinations. Then we can define

$$S_T = \sum_{t=1}^{T} W_{j,t}^{(i)} W_{j,t}^{(i')} - T\gamma^{\iota},$$

where therefore $\gamma^{\iota} \equiv \gamma_j^{(i,i')}$ thereby defining the equivalent of $T_n$ for the present paper.

With the above discussion in mind, the steps of the proof of [1, Theorem 7] can be translated directly to the setting of this paper. For this reason, we will only focus on a specific step which needs to be slightly adapted in order to understand where our conditions are needed. This step consists in proving

$$\left\| \mathcal{P}_0 \left( W_{j,t}^{(i)} W_{j,t}^{(i')} \right) \right\|_2 < \infty,$$

whose equivalent is proven for $\| \mathcal{P}_0 \left( X_i X_{i+j} \right) \|_2$ in Equation (64) in [1]. Based on the latter, it is easy to get to the following inequality

$$\left\| \mathcal{P}_0 \left( W_{j,t}^{(i)} W_{j,t}^{(i')} \right) \right\|_2 \leq \left\| W_{j,t}^{(i)} \right\|_4 \left\| W_{j,t}^{(i')} - W_{j,t}^{(i')\star} \right\|_4 + \left\| W_{j,t}^{(i)\star} - W_{j,t}^{(i)} \right\|_4 \left\| W_{j,t}^{(i')\star} \right\|_4.$$

We can see that the term on the right-hand side of the inequality can be re-expressed in terms

of the coefficients $c_{j,l}$ and the first-order differences $\Delta_t$ thereby obtaining the term

$$\left\|\sum_{l=0}^{2^j-2} c_{j,l}\Delta_{t-l}^{(i)}\right\|_4 \left\|\sum_{l=0}^{2^j-2} c_{j,l}\left(\Delta_{t-l}^{(i')}-\Delta_{t-l}^{(i')\star}\right)\right\|_4 + \left\|\sum_{l=0}^{2^j-2} c_{j,l}\left(\Delta_{t-l}^{(i)}-\Delta_{t-l}^{(i)\star}\right)\right\|_4 \left\|\sum_{l=0}^{2^j-2} c_{j,l}\Delta_{t-l}^{(i')}\right\|_4,$$

which we denote as $A$ (to be concise). Since $\max_{j,l} c_{j,l} = 1/2$ and the maximum length of the filter coefficients is $2^J - 1$, we have that

$$A \leq 2\,\frac{2^J-1}{4}\,\max_i\left\{\left\|\Delta_t^{(i)}\right\|_4\right\}\,\max_i\left\{\left\|\sum_{l=0}^{2^j-2}(\Delta_{t-l}^{(i)}-\Delta_{t-l}^{(i)\star})\right\|_4\right\}$$

$$\leq \frac{2^J-1}{2}\,\max_i\left\{\left\|\Delta_t^{(i)}\right\|_4\right\}\,\max_i\left\{\sum_{l=0}^{2^j-2}\left\|\Delta_{t-l}^{(i)}-\Delta_{t-l}^{(i)\star}\right\|_4\right\}$$

$$\leq \frac{2^J-1}{2}\,\max_i\left\{\left\|\Delta_t^{(i)}\right\|_4\right\}\,\max_i\left\{\left\|\Delta_t^{(i)}-\Delta_t^{(i)\star}\right\|_4\right\}.$$

Based on the fact that $J$ is fixed, we can now use Conditions **(C3)** and **(C4)** to show that

$$\left\|\mathcal{P}_0\left(W_{j,t}^{(i)}W_{j,t}^{(i')}\right)\right\|_2 < \infty.$$

Following the proof of [1, Theorem 7], we have that

$$\|S_T\|_2 = \sqrt{T}\,\|\tilde{\gamma}^\iota - \gamma^\iota\|_2 \leq T\,\left\|\mathcal{P}_0\left(W_{j,t}^{(i)}W_{j,t}^{(i')}\right)\right\|_2.$$

Using this result and the fact that $I$ is fixed, as for the proof of [1, Theorem 7], we satisfy the conditions for weak convergence in [2] and apply the Cramer-Wold device to obtain the result of asymptotic normality for the vector $\tilde{\boldsymbol{\nu}}$ thus concluding the proof of Lemma **A.1**. $\qquad\square$

We now need to show that the quantity $\sqrt{T}(\hat{\boldsymbol{\nu}} - \tilde{\boldsymbol{\nu}})$ converges to zero in order to transfer the asymptotic normality of $\tilde{\boldsymbol{\nu}}$ to $\hat{\boldsymbol{\nu}}$. For this reason, letting $|\boldsymbol{x}|_2 = \sqrt{\boldsymbol{x}^T\boldsymbol{x}}$ represent the L2-norm of the vector $\boldsymbol{x}$, the second step of the proof is given by the following lemma.

LEMMA A.2: *Under Conditions **(C2)**-**(C4)**, and with both $I \in \mathbb{N}$ and $J \in \mathbb{N}$, we have that*

$$\sqrt{T}\,|\hat{\boldsymbol{\nu}} - \tilde{\boldsymbol{\nu}}|_2 = o_p(1).$$

*Proof of Lemma A.2.* Recall that $M_j = T - 2^j + 1$ is the number of wavelet coefficients generated

at level $j$. By the definitions of $\tilde{\gamma}^\iota$ and $\hat{\gamma}^\iota$, we have for all $\iota$

$$\sqrt{T}\left(\tilde{\gamma}^\iota - \hat{\gamma}^\iota\right) = \frac{1}{\sqrt{T}}\sum_{t=1}^{T} W_{j,t}^{(i)} W_{j,t}^{(i')} - \frac{\sqrt{T}}{M_j}\sum_{t=L_j}^{T} W_{j,t}^{(i)} W_{j,t}^{(i')}.$$

We add and subtract $\sqrt{T}\gamma^\iota = \sqrt{T}\,\mathbb{E}\left[W_{j,t}^{(i)} W_{j,t}^{(i')}\right]$ in the above equation to obtain

$$\frac{1}{\sqrt{T}}\sum_{t=1}^{T} W_{j,t}^{(i)} W_{j,t}^{(i')} - \frac{\sqrt{T}}{M_j}\sum_{t=L_j}^{T} W_{j,t}^{(i)} W_{j,t}^{(i')} - \frac{T}{\sqrt{T}}\gamma^\iota + \frac{\sqrt{T}}{M_j}M_j\gamma^\iota$$

$$= \frac{1}{\sqrt{T}}\sum_{t=1}^{T}\left(W_{j,t}^{(i)} W_{j,t}^{(i')} - \gamma^\iota\right) - \frac{\sqrt{T}}{M_j}\sum_{t=L_j}^{T}\left(W_{j,t}^{(i)} W_{j,t}^{(i')} - \gamma^\iota\right)$$

$$= \underbrace{\frac{M_j - T}{M_j\sqrt{T}}\sum_{t=1}^{T}\left(W_{j,t}^{(i)} W_{j,t}^{(i')} - \gamma^\iota\right)}_{A} \underbrace{- \frac{\sqrt{T}}{M_j}\sum_{t=1}^{2^j-1}\left(W_{j,t}^{(i)} W_{j,t}^{(i')} - \gamma^\iota\right)}_{B}.$$

We can now separately study the terms $A$ and $B$. Starting from term $A$, we have that

$$\frac{M_j - T}{M_j} = \mathcal{O}\left(\frac{1}{T}\right)$$

and, based on Lemma A.1, we also have

$$\frac{1}{\sqrt{T}}\sum_{t=1}^{T}\left(W_{j,t}^{(i)} W_{j,t}^{(i')} - \gamma^\iota\right) = \mathcal{O}_p(1).$$

Therefore we finally have

$$A = \mathcal{O}_p\left(\frac{1}{T}\right) = o_p(1).$$

Turning to term $B$, recall that it is defined as

$$B = -\frac{\sqrt{T}}{M_j}\sum_{t=1}^{2^j-1}\left(W_{j,t}^{(i)} W_{j,t}^{(i')} - \gamma^\iota\right).$$

where, based on the fact that $J$ is fixed (finite), and Conditons (C3) and (C4), we have that for any $\epsilon > 0$

$$\lim_{T\to\infty}\mathbb{P}\left(|B| > \epsilon\right) \leq \lim_{T\to\infty}\frac{\|B\|_2^2}{\epsilon^2} \leq \lim_{T\to\infty}\frac{T(2^j-1)\left\|\mathcal{P}_0\left(W_{j,t}^{(i)} W_{j,t}^{(i')}\right)\right\|_2^2}{T^2\epsilon^2} = 0.$$

Therefore, we have

$$B = o_p(1).$$

Based on these terms, we can now study the quantity

$$
\sqrt{T} \, |\hat{\boldsymbol{\nu}} - \tilde{\boldsymbol{\nu}}|_2 \leq \sum_{\substack{j=1,\ldots,J \\ 1 \leq i \leq i' \leq I}} \sqrt{T} \left| \tilde{\gamma}_j^{(i,i')} - \hat{\gamma}_j^{(i,i')} \right|
$$

$$
\leq \sum_{\substack{j=1,\ldots,J \\ 1 \leq i \leq i' \leq I}} \left( |A| + |B| \right) = o_p(1),
$$

thus concluding the proof of Lemma A.2. $\qquad\square$

We can now combine the results of Lemmas A.1 and A.2 based on the decomposition in (1). Indeed, using Slutsky's theorem, the asymptotic normality of $\hat{\boldsymbol{\nu}}$ directly follows, thus concluding the proof of Theorem 1.

*B. Proofs of Identifiability*

In this section, we prove Proposition 1, Lemma 1 and Proposition 2 which discuss the identifiability of Model **(M1)** and Model **(M2)** respectively. These results can be proved by using [3, Theorem 2] which states that, for a function $f : \mathbb{R}^p \to \mathbb{R}^p$, its injectivity is guaranteed by the following conditions:

**(I1)** For $\boldsymbol{\theta} \in \mathbb{R}^p$, the map $f(\boldsymbol{\theta})$ is twice continuously differentiable in $\mathbb{R}^p$.

**(I2)** $|f(\boldsymbol{\theta})|_2 \to \infty$ whenever $|\boldsymbol{\theta}|_2 \to \infty$.

**(I3)** For $\boldsymbol{\theta} \in \mathbb{R}^p$, the Jacobian of $f(\boldsymbol{\theta})$ is nonnegative.

**(I4)** For every $\boldsymbol{C} \in \mathbb{R}^p$ the equation $f(\boldsymbol{\theta}) = \boldsymbol{C}$ has countably many (possibly zero) solutions in $\mathbb{R}^p$.

We will now prove these conditions in order to deliver the identifiability results claimed for the class of models **(M1)** and **(M2)**.

*Proof of Proposition 1.* Let us consider the theoretical WCCV vector for Model **(M1)** defined as

$$\boldsymbol{\nu}(\boldsymbol{\theta}) \equiv \left[ \gamma_j^{(i,i')} \right]_{\substack{j=1,\dots,J \\ 1 \le i \le i' \le I}},$$

where

$$\gamma_j^{(i,i')} = \begin{cases} \dfrac{\sigma^{(i,i')}}{2^j} + \dfrac{2^{2j-1}+1}{6}\lambda^{(i,i')} + \dfrac{6}{2^{2j}}Q_i^2 + \dfrac{2^{2j}}{16}\omega_i & \text{if } i = i', \\[2ex] \dfrac{\sigma^{(i,i')}}{2^j} + \dfrac{2^{2j-1}+1}{6}\lambda^{(i,i')} & \text{else.} \end{cases} \tag{2}$$

Indeed, when $i \ne i'$, the WCCV for **(T3)** and **(T4)** is null by definition. Now let

$$\boldsymbol{\theta} \equiv \left[ \boldsymbol{\sigma}^\mathsf{T}, \;\; \boldsymbol{\lambda}^\mathsf{T}, \;\; \boldsymbol{Q}^\mathsf{T}, \;\; \boldsymbol{\omega}^\mathsf{T} \right]^\mathsf{T},$$

where

$$\boldsymbol{\sigma} \equiv \left[ \sigma^{(1,1)}, \;\; \sigma^{(1,2)}, \;\; \dots, \;\; \sigma^{(I,I)} \right]^\mathsf{T},$$

$$\boldsymbol{\lambda} \equiv \left[ \lambda^{(1,1)}, \;\; \lambda^{(1,2)}, \;\; \dots, \;\; \lambda^{(I,I)} \right]^\mathsf{T},$$

$$\boldsymbol{Q} \equiv \left[ Q_1^2, \;\; Q_2^2, \;\; \dots, \;\; Q_I^2 \right]^\mathsf{T},$$

and

$$\boldsymbol{\omega} \equiv \begin{bmatrix} \omega_1, & \omega_2, & \ldots, & \omega_I \end{bmatrix}^{\mathsf{T}}.$$

The vector $\boldsymbol{\theta}$ therefore represents the parameter vector containing all the parameters for each model contributing to the $I$-dimensional multivariate time series. In order to use [3, Theorem 2], we need to rearrange this parameter vector in order to respect Condition **(I1)** which requires that the parameter vector $\boldsymbol{\theta}$ be of the same dimension as the vector $\boldsymbol{\nu}(\boldsymbol{\theta})$. To do so, and also for the clarity of the derivations that will follow, we rearrange $\boldsymbol{\theta}$ as $\boldsymbol{\theta} = \begin{bmatrix} \boldsymbol{\theta}_1^{\mathsf{T}}, & \boldsymbol{\theta}_2^{\mathsf{T}} \end{bmatrix}^{\mathsf{T}}$ where $\boldsymbol{\theta}_1$ collects all the parameters that strictly represent individual time series, i.e. for $i = i'$ we have

$$\boldsymbol{\theta}_1 = \begin{bmatrix} \sigma^{(1,1)}, & \lambda^{(1,1)}, & Q_1^2, & \omega_1, & \ldots, & \sigma^{(I,I)}, & \lambda^{(I,I)}, & Q_I^2, & \omega_I \end{bmatrix}^{\mathsf{T}},$$

while $\boldsymbol{\theta}_2$ collects all those parameters that describe the dependence between signals, i.e. for $i \neq i'$ we have

$$\boldsymbol{\theta}_2 = \begin{bmatrix} \sigma^{(1,2)}, & \lambda^{(1,2)}, & \ldots, & \sigma^{(I-1,I)}, & \lambda^{(I-1,I)} \end{bmatrix}^{\mathsf{T}}.$$

In the same logic, we also rearrange $\boldsymbol{\nu}(\boldsymbol{\theta})$ as $\boldsymbol{\nu}(\boldsymbol{\theta}) = \begin{bmatrix} \boldsymbol{\nu}_1(\boldsymbol{\theta})^{\mathsf{T}}, & \boldsymbol{\nu}_2(\boldsymbol{\theta})^{\mathsf{T}}, & \boldsymbol{\nu}_3(\boldsymbol{\theta})^{\mathsf{T}} \end{bmatrix}^{\mathsf{T}}$, where for WCCV with $i = i'$ and $j = 1, 2, 3, 4$, we have

$$\boldsymbol{\nu}_1(\boldsymbol{\theta}) = \boldsymbol{\nu}_1(\boldsymbol{\theta}_1) = \begin{bmatrix} \gamma_1^{(1,1)}, & \ldots, & \gamma_4^{(1,1)}, & \ldots, & \gamma_1^{(I,I)}, & \ldots, & \gamma_4^{(I,I)} \end{bmatrix}^{\mathsf{T}},$$

while for WCCV with $i \neq i'$ and $j = 1, 2$, we have

$$\boldsymbol{\nu}_2(\boldsymbol{\theta}) = \boldsymbol{\nu}_2(\boldsymbol{\theta}_2) \equiv \begin{bmatrix} \gamma_1^{(1,2)}, & \gamma_2^{(1,2)}, \cdots, & \gamma_1^{(I-1,I)}, & \gamma_2^{(I-1,I)} \end{bmatrix}^{\mathsf{T}}.$$

Using these definitions we can finally define $\boldsymbol{\nu}^{\star}(\boldsymbol{\theta}) \equiv \begin{bmatrix} \boldsymbol{\nu}_1(\boldsymbol{\theta}_1)^{\mathsf{T}}, & \boldsymbol{\nu}_2(\boldsymbol{\theta}_2)^{\mathsf{T}} \end{bmatrix}^{\mathsf{T}}$ by discarding $\boldsymbol{\nu}_3(\boldsymbol{\theta})$ which ensures that $\boldsymbol{\theta}$ and $\boldsymbol{\nu}^{\star}(\boldsymbol{\theta})$ share the same dimension. Indeed, proving identifiability for $\boldsymbol{\nu}^{\star}(\boldsymbol{\theta})$ implies identifiability for the entire vector $\boldsymbol{\nu}(\boldsymbol{\theta})$.

An additional requirement of Condition **(I1)** is that $\boldsymbol{\theta}_1$ is in $\mathbb{R}^{4I}$, which is not true since all the parameters in $\boldsymbol{\theta}_1$ are variances. In fact, we have that

$$\boldsymbol{\theta}_1 \in \mathbb{R}_+^{4I}.$$

To solve this, we introduce an injective element-wise operator $g$ and its inverse $g^{-1}$ which allows us to define

$$\boldsymbol{\theta}_1^{\star} \equiv g(\boldsymbol{\theta}_1),$$

where

$$g(\boldsymbol{\theta}_1) \equiv \Big[\log(\sigma^{(1,1)}), \ \log(\lambda^{(1,1)}), \ \log(Q_1^2), \ \log(\omega_1), \ \ldots, \ \log(\sigma^{(I,I)}), \ \log(\lambda^{(I,I)}), \ \log(Q_I^2), \ \log(\omega_I)\Big]^{\mathsf{T}}$$

$$\equiv \Big[\sigma^{(1,1)\star}, \ \lambda^{(1,1)\star}, \ Q_1^{2\star}, \ \omega_1^{\star}, \ \ldots, \ \sigma^{(I,I)\star}, \ \lambda^{(I,I)\star}, \ Q_I^{2\star}, \ \omega_I^{\star}\Big]^{\mathsf{T}}.$$

Consequently we have

$$\boldsymbol{\theta}_1 \equiv g^{-1}(\boldsymbol{\theta}_1^{\star}),$$

where

$$g^{-1}(\boldsymbol{\theta}_1^{\star}) \equiv \Big[\exp(\sigma^{(1,1)\star}), \ \exp(\lambda^{(1,1)\star}), \ \ldots, \ \exp(Q_I^{2\star}), \ \exp(\omega_I^{\star})\Big]^{\mathsf{T}}.$$

By construction, we have $\boldsymbol{\theta}_1^{\star} \in \mathbb{R}^{4I}$ which brings us to define

$$\boldsymbol{\theta}^{\star} \equiv \Big[\boldsymbol{\theta}_1^{\star T}, \ \boldsymbol{\theta}_2^{\mathsf{T}}\Big]^{\mathsf{T}} \in \mathbb{R}^{4I + I(I-1)}.$$

Hence, our goal is now to show that $\boldsymbol{\nu}^{\star}(\boldsymbol{\theta}^{\star}) = \Big[\boldsymbol{\nu}_1(g^{-1}(\boldsymbol{\theta}_1^{\star}))^{\mathsf{T}}, \ \boldsymbol{\nu}_2(\boldsymbol{\theta}_2)^{\mathsf{T}}\Big]^{\mathsf{T}}$ is an injective function of $\boldsymbol{\theta}^{\star}$. This is equivalent to showing that $\boldsymbol{\nu}^{\star}(\boldsymbol{\theta})$ is an injective function of $\boldsymbol{\theta}$ as well as being equivalent to showing that $\boldsymbol{\nu}(\boldsymbol{\theta})$ is an injective function of $\boldsymbol{\theta}$.

Having structured the problem to respect the requirements of Condition **(I1)**, we now set out to verify Conditions **(I1)**-**(I4)**. To start, it is trivial to see that Condition **(I2)** is satisfied, i.e.

$$|\boldsymbol{\nu}^{\star}(\boldsymbol{\theta}^{\star})|_2 \to \infty \ \text{ whenever } |\boldsymbol{\theta}^{\star}|_2 \to \infty. \tag{3}$$

Having verified this, we now consider the derivatives with respect to $\sigma^{(l,m)}$, $\lambda^{(l,m)}$, $Q_l^2$ and $\omega_l$ to verify Condition **(I1)**. These derivatives are respectively

$$a_{j,(1)}^{(i,i')}(l,m) \equiv \frac{\partial}{\partial \sigma^{(l,m)\star}} \gamma_j^{(i,i')}$$

$$= \begin{cases} \dfrac{1}{2^j} \exp(\sigma^{(l,m)\star}) & \text{if } i = l = i' = m, \\[2ex] \dfrac{1}{2^j} & \text{if } i = l \neq i' = m, \\[2ex] 0 & \text{else,} \end{cases}$$

$$a_{j,(2)}^{(i,i')}(l,m) \equiv \frac{\partial}{\partial \lambda^{(l,m)\star}} \gamma_j^{(i,i')}$$

$$= \begin{cases} \dfrac{2^{2j-1}+1}{2^j 6} \exp(\lambda^{(l,m)\star}) & \text{if } i=l=i'=m, \\[2ex] \dfrac{2^{2j-1}+1}{2^j 6} & \text{if } i=l \neq i'=m, \\[2ex] 0 & \text{else}, \end{cases}$$

$$a_{j,(3)}^{(i,i')}(l,l) \equiv \frac{\partial}{\partial Q_l^{2\star}} \gamma_j^{(i,i')}$$

$$= \begin{cases} \dfrac{6}{2^{2j}} \exp(Q_l^{2\star}) & \text{if } i=i'=l, \\[2ex] 0 & \text{else}, \end{cases}$$

$$a_{j,(4)}^{(i,i')}(l,l) \equiv \frac{\partial}{\partial \omega_l^{\star}} \gamma_j^{(i,i')}$$

$$= \begin{cases} \dfrac{2^{2j}}{16} \exp(\omega_l^{\star}) & \text{if } i=i'=l, \\[2ex] 0 & \text{else}. \end{cases}$$

By simple verification based on the expressions above, we have that the map $\boldsymbol{\nu}^{\star}(\boldsymbol{\theta}^{\star})$ is in $\mathcal{C}^2(\mathbb{R}^{4I+I(I-1)})$ which consequently verifies Condition **(I1)**.

To verify the last two conditions (i.e. **(I3)** and **(I4)**), we study the Jacobian matrix $\boldsymbol{A}^{\star}$ of $\boldsymbol{\nu}^{\star}(\boldsymbol{\theta}^{\star}) \equiv \left[\boldsymbol{\nu}_1(g^{-1}(\boldsymbol{\theta}_1^{\star}))^{\mathsf{T}}, \ \boldsymbol{\nu}_2(\boldsymbol{\theta}_2)^{\mathsf{T}}\right]^{\mathsf{T}}$ with respect to $\boldsymbol{\theta}^{\star}$ which is given by

$$\boldsymbol{A}^{\star} \equiv \begin{bmatrix} \frac{\partial}{\partial \boldsymbol{\theta}_1^{\star}} \boldsymbol{\nu}_1(g^{-1}(\boldsymbol{\theta}_1^{\star})) & \frac{\partial}{\partial \boldsymbol{\theta}_2} \boldsymbol{\nu}_1(g^{-1}(\boldsymbol{\theta}_1^{\star})) \\[1ex] \frac{\partial}{\partial \boldsymbol{\theta}_1^{\star}} \boldsymbol{\nu}_2(\boldsymbol{\theta}_2) & \frac{\partial}{\partial \boldsymbol{\theta}_2} \boldsymbol{\nu}_2(\boldsymbol{\theta}_2) \end{bmatrix}$$

$$= \begin{bmatrix} \boldsymbol{A}_1 & \boldsymbol{0}_{4I \times I(I-1)} \\[1ex] \boldsymbol{0}_{I(I-1) \times 4I} & \boldsymbol{A}_2 \end{bmatrix},$$

where

$$\boldsymbol{A}_1 = \begin{bmatrix} a_{1,(1)}^{(1,1)}(1,1) & a_{1,(2)}^{(1,1)}(1,1) & \cdots & a_{1,(3)}^{(1,1)}(I,I) & a_{1,(4)}^{(1,1)}(I,I) \\ a_{2,(1)}^{(1,1)}(1,1) & a_{2,(2)}^{(1,1)}(1,1) & \cdots & a_{2,(3)}^{(1,1)}(I,I) & a_{2,(4)}^{(1,1)}(I,I) \\ a_{3,(1)}^{(1,1)}(1,1) & a_{3,(2)}^{(1,1)}(1,1) & \cdots & a_{3,(3)}^{(1,1)}(I,I) & a_{3,(4)}^{(1,1)}(I,I) \\ a_{4,(1)}^{(1,1)}(1,1) & a_{4,(2)}^{(1,1)}(1,1) & \cdots & a_{4,(3)}^{(1,1)}(I,I) & a_{4,(4)}^{(1,1)}(I,I) \\ \vdots & \vdots & \ddots & \vdots & \vdots \\ a_{3,(1)}^{(I,I)}(1,1) & a_{3,(2)}^{(I,I)}(1,1) & \cdots & a_{3,(3)}^{(I,I)}(I,I) & a_{3,(4)}^{(I,I)}(I,I) \\ a_{4,(1)}^{(I,I)}(1,1) & a_{4,(2)}^{(I,I)}(1,1) & \cdots & a_{4,(3)}^{(I,I)}(I,I) & a_{4,(4)}^{(I,I)}(I,I) \end{bmatrix}$$

and

$$\begin{aligned} \boldsymbol{A}_2 &= \begin{bmatrix} a_{1,(1)}^{(1,2)}(1,2) & a_{1,(2)}^{(1,2)}(1,2) & \cdots & a_{1,(1)}^{(1,2)}(I-1,I) & a_{1,(2)}^{(1,2)}(I-1,I) \\ a_{2,(1)}^{(1,2)}(1,2) & a_{2,(2)}^{(1,2)}(1,2) & \cdots & a_{2,(1)}^{(1,2)}(I-1,I) & a_{2,(2)}^{(1,2)}(I-1,I) \\ a_{1,(1)}^{(1,3)}(1,2) & a_{1,(2)}^{(1,3)}(1,2) & \cdots & a_{1,(1)}^{(1,3)}(I-1,I) & a_{1,(2)}^{(1,3)}(I-1,I) \\ a_{2,(1)}^{(1,3)}(1,2) & a_{2,(2)}^{(1,3)}(1,2) & \cdots & a_{2,(1)}^{(1,3)}(I-1,I) & a_{2,(2)}^{(1,3)}(I-1,I) \\ \vdots & \vdots & \ddots & \vdots & \vdots \\ a_{1,(1)}^{(I-1,I)}(1,2) & a_{1,(2)}^{(I-1,I)}(1,2) & \cdots & a_{1,(1)}^{(I-1,I)}(I-1,I) & a_{1,(2)}^{(I-1,I)}(I-1,I) \\ a_{2,(1)}^{(I-1,I)}(1,2) & a_{2,(2)}^{(I-1,I)}(1,2) & \cdots & a_{2,(1)}^{(I-1,I)}(I-1,I) & a_{2,(2)}^{(I-1,I)}(I-1,I) \end{bmatrix} \\[2mm] &= \begin{bmatrix} \frac{1}{2} & \frac{1}{4} & 0 & 0 & \cdots & 0 & 0 \\ \frac{1}{4} & \frac{3}{8} & 0 & 0 & \cdots & 0 & 0 \\ 0 & 0 & \frac{1}{2} & \frac{1}{4} & \cdots & 0 & 0 \\ 0 & 0 & \frac{1}{4} & \frac{3}{8} & \cdots & 0 & 0 \\ \vdots & \vdots & \vdots & \vdots & \ddots & \vdots & \vdots \\ 0 & 0 & 0 & 0 & \cdots & \frac{1}{2} & \frac{1}{4} \\ 0 & 0 & 0 & 0 & \cdots & \frac{1}{4} & \frac{3}{8} \end{bmatrix}. \end{aligned}$$

By straighforward computation, we have the determinants $|\boldsymbol{A}_1| = \left(\frac{2205}{4096}\right)^I \exp\left(\sum_{i=1}^I \sigma^{(i,i)\star}\right) > 0$ and $|\boldsymbol{A}_2| = \left(\frac{1}{8}\right)^{\frac{I(I-1)}{2}} > 0$ which constitute the diagonal matrices of a block diagonal matrix whose determinant is the product of their determinants. Thus Condition **(I3)** is verified as well as Condition **(I4)** since we have $|\boldsymbol{A}^\star| > 0$. Consequently we have that $\boldsymbol{\nu}^\star(\boldsymbol{\theta}^\star)$ is an injective function of $\boldsymbol{\theta}^\star$ which implies identifiability of Model **(M1)** through the WCCV vector. $\square$

In order to prove Lemma **1** we need to show three auxilary results, namely Lemma **A.3**, **A.4** and **A.5**, which provide the closed form of the theoretical WCCV of Models **(T1)** (WN), **(T2)**

(RW) and **(T5)** (AR(1)) (the others are not considered since their WCCV is null by definition when $i \neq i'$).

Lemma A.3: *For model* **(T1)**, *i.e.* $\boldsymbol{X}_{1,t} \overset{iid}{\sim} \mathcal{F}(\boldsymbol{0}, \boldsymbol{\Sigma})$, *the theoretical WCCV (see Equation (4)) is given by*

$$\gamma_j^{(i,i')} = \frac{\sigma^{(i,i')}}{2^j}, \text{ for } 1 \leq i, i' \leq I \text{ and } 1 \leq j \leq J,$$

*where* $\sigma^{(i,i')}$ *is the* $(i,i')^{th}$ *element of* $\boldsymbol{\Sigma}$.

*Proof of Lemma A.3.* By the definition of the WCCV, and since $\boldsymbol{X}_{1,t}$ is has null expectation, we have

$$\gamma_j^{(i,i')} = \mathbb{E}\left[\left(\sum_{l=0}^{L_j-1} h_{j,l}\, X_{1,t-l}^{(i)}\right)\left(\sum_{l=0}^{L_j-1} h_{j,l}\, X_{1,t-l}^{(i')}\right)\right]$$

$$= \mathbb{E}\left[\sum_{l=0}^{L_j-1} h_{j,l}^2\, X_{1,t-l}^{(i)} X_{1,t-l}^{(i')}\right]$$

$$= \sum_{l=0}^{L_j-1} h_{j,l}^2\, \mathbb{E}\left[X_{1,t-l}^{(i)} X_{1,t-l}^{(i')}\right] = \sum_{l=0}^{L_j-1} h_{j,l}^2\, \sigma^{(i,i')}$$

$$= \frac{\sigma^{(i,i')}}{2^j}.$$

$\square$

Lemma A.4: *For model* **(T2)**, *i.e.* $\boldsymbol{X}_{2,t} = \boldsymbol{X}_{2,t-1} + \boldsymbol{\iota}_t$, *where* $\boldsymbol{\iota}_t \overset{iid}{\sim} \mathcal{F}(\boldsymbol{0}, \boldsymbol{\Lambda})$, *the theoretical WCCV is given by*

$$\gamma_j^{(i,i')} = \frac{2^{2j}+2}{2^j\, 12} \lambda^{(i,i')}, \text{ for } 1 \leq i, i' \leq I \text{ and } 1 \leq j \leq J,$$

*where* $\lambda^{(i,i')}$ *is the* $(i,i')^{th}$ *element of* $\boldsymbol{\Lambda}$.

*Proof of Lemma A.4.* By applying Haar wavelet filter to each random walk process $i = 1, ..., I$, each wavelet coefficient $W_{j,t}^{(i)}$ is equal to $\frac{1}{2^j}$ times the sum of all the i.i.d. random variables listed in the following matrix

$$\begin{bmatrix} \iota_{t-2^j+2}^{(i)} & \iota_{t-2^j+3}^{(i)} & \cdots & \iota_{t-2^{j-1}+1}^{(i)} \\ \iota_{t-2^j+3}^{(i)} & \iota_{t-2^j+4}^{(i)} & \cdots & \iota_{t-2^{j-1}+2}^{(i)} \\ \vdots & \vdots & \ddots & \cdots \\ \iota_{t-2^{j-1}+1}^{(i)} & \iota_{t-2^{j-1}+2}^{(i)} & \cdots & \iota_t^{(i)} \end{bmatrix},$$

i.e.

$$W_{j,t}^{(i)} = \frac{1}{2^j} \left( \sum_{i=1}^{2^{j-1}} i \, \iota_{t-2^j+2+i} + \mathbb{1}_{j\geq 2} \sum_{i=2^{j-1}+1}^{2^j-1} \left( 2^j - i \right) \iota_{t-2^j+2+i} \right),$$

where $\mathbb{1}.$ represents the indicator function. Then by definition, Eq. (4) can be simplified as

$$\gamma_j^{(i,i')} = \frac{1}{2^{2j}} \left( \left( 2^{j-1} \right)^2 \lambda^{(i,i')} + 2 \sum_{i=1}^{2^{j-1}-1} i^2 \lambda^{(i,i')} \right)$$

$$= \frac{1}{2^{2j}} \left( 2^{2j-2} + \frac{\left( 2^{j-1} - 1 \right) 2^{j-1} \left( 2^j - 1 \right)}{3} \right) \lambda^{(i,i')}$$

$$= \frac{2^{2j}+2}{2^j \, 12} \lambda^{(i,i')}.$$

<div align="right">□</div>

LEMMA A.5: *For model (T5), i.e. $\mathbf{X}_{k,t} = \mathbf{\Phi}\mathbf{X}_{k,t-1} + \boldsymbol{\varepsilon}_t$, with $k \geq 5$ where $\mathbf{\Phi}$ is a diagonal matrix with diagonal elements $\phi_k^{(i)}$, $0 < |\phi_k^{(i)}| < 1$, for $i = 1, \ldots, I$ and $\boldsymbol{\varepsilon}_t \overset{iid}{\sim} \mathcal{F}(\mathbf{0}, \mathbf{Z}_k)$. the theoretical WCCV is given by*

$$\gamma_j^{(i,i')} = \left\{ 2^j + 2 \left[ \frac{\phi_k^{(i)}}{1-\phi_k^{(i)}} \left( 2^{j-1} - 1 \right) - \left( \frac{\phi_k^{(i)}}{1-\phi_k^{(i)}} \right)^2 \left( 1 - \left( \phi_k^{(i)} \right)^{2^{j-1}-1} \right) \right] \right.$$

$$+ 2 \left[ \frac{\phi_k^{(i')}}{1-\phi_k^{(i')}} \left( 2^{j-1} - 1 \right) - \left( \frac{\phi_k^{(i')}}{1-\phi_k^{(i')}} \right)^2 \left( 1 - \left( \phi_k^{(i')} \right)^{2^{j-1}-1} \right) \right]$$

$$\left. - \left[ \frac{1 - \left( \phi_k^{(i)} \right)^{2^{j-1}}}{1-\phi_k^{(i)}} \right]^2 \phi_k^{(i)} - \left[ \frac{1 - \left( \phi_k^{(i')} \right)^{2^{j-1}}}{1-\phi_k^{(i')}} \right]^2 \phi_k^{(i')} \right\} \frac{z_k^{(i,i')}}{4^j \left( 1 - \phi_k^{(i)} \, \phi_k^{(i')} \right)},$$

*and the Cross-Spectrum is given by*

$$\boldsymbol{S_\theta}^{(i,i')}(f) = \frac{e^{i2\pi f} z_k^{(i,i')}}{\left( \phi_k^{(i)} e^{i2\pi f} - 1 \right) \left( \phi_k^{(i')} - e^{i2\pi f} \right)},$$

*where $z_k^{(i,i')}$ is the $(i, i')^{th}$ element of $\mathbf{Z}_k$.*

*Proof of Lemma A.5.* Since the model (T5) is jointly stationary, the cross-covariance with respect to directed lag $l$ is

$$C_{0,k}^{(i,i')}(\boldsymbol{\theta}) = \frac{z_k^{(i,i')}}{1-\phi_k^{(i)} \, \phi_k^{(i')}},$$

when $l = 0$. When the lag $l \neq 0$, we have that

$$C_{l,k}^{(i,i')}(\boldsymbol{\theta}) = \left(\phi_k^{(i)}\right)^{-l} C_{0,k}^{(i,i')}(\boldsymbol{\theta}),$$

when $l < 0$ and

$$C_{l,k}^{(i,i')}(\boldsymbol{\theta}) = \left(\phi_k^{(i')}\right)^{l} C_{0,k}^{(i,i')}(\boldsymbol{\theta}),$$

when $l > 0$. Thus we have

$$
\gamma_j^{(i,i')} = \begin{bmatrix} h_{j,0} \\ h_{j,1} \\ \vdots \\ h_{j,L_j-1} \end{bmatrix}^{\mathsf{T}} \begin{bmatrix} C_{0,k}^{(i,i')}(\boldsymbol{\theta}) & \cdots & C_{L_j-1,k}^{(i,i')}(\boldsymbol{\theta}) \\ C_{-1,k}^{(i,i')}(\boldsymbol{\theta}) & \cdots & C_{L_j-2,k}^{(i,i')}(\boldsymbol{\theta}) \\ \vdots & \ddots & \vdots \\ C_{-L_j+1,k}^{(i,i')}(\boldsymbol{\theta}) & \cdots & C_{0,k}^{(i,i')}(\boldsymbol{\theta}) \end{bmatrix} \begin{bmatrix} h_{j,0} \\ h_{j,1} \\ \vdots \\ h_{j,L_j-1} \end{bmatrix}
$$

$$
= \begin{bmatrix} h_{j,0} \\ h_{j,1} \\ \vdots \\ h_{j,L_j-1} \end{bmatrix}^{\mathsf{T}} \begin{bmatrix} 1 & \cdots & \left(\phi_k^{(i')}\right)^{L_j-1} \\ \phi_k^{(i)} & \cdots & \left(\phi_k^{(i')}\right)^{L_j-2} \\ \vdots & \ddots & \vdots \\ \left(\phi_k^{(i)}\right)^{L_j-1} & \cdots & 1 \end{bmatrix} \begin{bmatrix} h_{j,0} \\ h_{j,1} \\ \vdots \\ h_{j,L_j-1} \end{bmatrix} C_{0,k}^{(i,i')}(\boldsymbol{\theta})
$$

$$
= \begin{bmatrix} h_{j,0} \\ h_{j,1} \\ \vdots \\ h_{j,L_j-1} \end{bmatrix}^{\mathsf{T}} \begin{bmatrix} 1 & \cdots & \left(\phi_k^{(i')}\right)^{L_j-1} \\ \phi_k^{(i)} & \cdots & \left(\phi_k^{(i')}\right)^{L_j-2} \\ \vdots & \ddots & \vdots \\ \left(\phi_k^{(i)}\right)^{L_j-1} & \cdots & 1 \end{bmatrix} \begin{bmatrix} h_{j,0} \\ h_{j,1} \\ \vdots \\ h_{j,L_j-1} \end{bmatrix} \frac{z_k^{(i,i')}}{1 - \phi_k^{(i)} \, \phi_k^{(i')}}
$$

$$
= \left\{ 2^j + 2 \left[ \frac{\phi_k^{(i)}}{1 - \phi_k^{(i)}} \left(2^{j-1} - 1\right) - \left(\frac{\phi_k^{(i)}}{1 - \phi_k^{(i)}}\right)^2 \left(1 - \left(\phi_k^{(i)}\right)^{2^{j-1}-1}\right) \right] \right.
$$

$$
+ 2 \left[ \frac{\phi_k^{(i')}}{1 - \phi_k^{(i')}} \left(2^{j-1} - 1\right) - \left(\frac{\phi_k^{(i')}}{1 - \phi_k^{(i')}}\right)^2 \left(1 - \left(\phi_k^{(i')}\right)^{2^{j-1}-1}\right) \right]
$$

$$
\left. - \left[ \frac{1 - \left(\phi_k^{(i)}\right)^{2^{j-1}}}{1 - \phi_k^{(i)}} \right]^2 \phi_k^{(i)} - \left[ \frac{1 - \left(\phi_k^{(i')}\right)^{2^{j-1}}}{1 - \phi_k^{(i')}} \right]^2 \phi_k^{(i')} \right\} \frac{z_k^{(i,i')}}{4^j \left(1 - \phi_k^{(i)} \, \phi_k^{(i')}\right)}.
$$

Moreover, the Cross-Spectrum for $\{X_{k,t}^{(i)}\}$ and $\{X_{k,t}^{(i')}\}$ is given by

$$
\begin{aligned}
\boldsymbol{S}_{\boldsymbol{\theta}}^{(i,i')}(f) &\equiv \sum_{l=-\infty}^{\infty} C_{l,k}^{(i,i')}(\boldsymbol{\theta})\, e^{-i2\pi l f} \\
&= \Bigg[ \sum_{l=-\infty}^{-1} \left( \phi_k^{(i)} \right)^{-l} e^{-i2\pi l f} \\
&\quad + \sum_{l=1}^{\infty} \left( \phi_k^{(i')} \right)^{l} e^{-i2\pi l f} + 1 \Bigg] \frac{z_k^{(i,i')}}{1 - \phi_k^{(i)}\,\phi_k^{(i')}} \\
&= \frac{e^{i2\pi f}\, z_k^{(i,i')}}{\left( \phi_k^{(i)} e^{i2\pi f} - 1 \right) \left( \phi_k^{(i')} - e^{i2\pi f} \right)}.
\end{aligned}
$$

$\square$

Having studied the closed forms of the WCCV for the above processes, we can now prove Lemma 1.

*Proof of Lemma 1.* For the sake of clarity, as for the previous proof we rearrange the order of elements in $\boldsymbol{\theta}$ as

$$
\boldsymbol{\theta} \equiv \left[ \boldsymbol{\theta}_1^{\mathsf{T}},\ \ \boldsymbol{\theta}_2^{\mathsf{T}} \right]^{\mathsf{T}},
$$

with

$$
\boldsymbol{\theta}_1 \equiv \left[ [\boldsymbol{\theta}^{(1,1)}]^{\mathsf{T}},\ \ [\boldsymbol{\theta}^{(2,2)}]^{\mathsf{T}},\ \ \cdots,\ \ [\boldsymbol{\theta}^{(I,I)}]^{\mathsf{T}} \right]^{\mathsf{T}},
$$

where each vector-element is given by

$$
\boldsymbol{\theta}^{(i,i)} \equiv \left[ \sigma^{(i,i)},\ \ Q_i^2,\ \ \phi_5^{(i)},\ \ \cdots,\ \ \phi_K^{(i)},\ \ z_5^{(i,i)},\ \ \cdots,\ \ z_K^{(i,i)} \right]^{\mathsf{T}} \in \mathbb{R}^{2K-6},
$$

for $i = 1, \cdots, I$ and

$$
\boldsymbol{\theta}_2 \equiv \left[ [\boldsymbol{\theta}^{(1,2)}]^{\mathsf{T}},\ \ [\boldsymbol{\theta}^{(1,3)}]^{\mathsf{T}},\ \ \cdots,\ \ [\boldsymbol{\theta}^{(I-1,I)}]^{\mathsf{T}} \right]^{\mathsf{T}},
$$

where each vector-element is given by

$$
\boldsymbol{\theta}^{(i,i')} \equiv \left[ \sigma^{(i,i')},\ \ z_5^{(i,i')},\ \ \cdots,\ \ z_K^{(i,i')} \right]^{\mathsf{T}} \in \mathbb{R}^{K-3},
$$

for $1 \leq i < i' \leq I$. In addition, we rearrange $\boldsymbol{C}(\boldsymbol{\theta})$ as follows

$$
\begin{aligned}
\boldsymbol{C}(\boldsymbol{\theta}) &\equiv \left[ C_l^{(i,i')}(\boldsymbol{\theta}) \right]_{\substack{l \in \mathbb{Z} \\ 1 \leq i \leq i' \leq I}}^{\mathsf{T}} \\
&= \left[ [C_l^{(1,1)}(\boldsymbol{\theta})]_{l \in \mathbb{Z}}^{\mathsf{T}}, \quad \cdots, \quad [C_l^{(I,I)}(\boldsymbol{\theta})]_{l \in \mathbb{Z}}^{\mathsf{T}}, \quad [C_l^{(1,2)}(\boldsymbol{\theta})]_{l \in \mathbb{Z}}^{\mathsf{T}}, \quad \cdots, \quad [C_l^{(I-1,I)}(\boldsymbol{\theta})]_{l \in \mathbb{Z}}^{\mathsf{T}} \right]^{\mathsf{T}}.
\end{aligned}
$$

For each $i = 1, \cdots, I$, the function $C_l^{(i,i)}$ only depends on the subvector $\boldsymbol{\theta}^{(i,i)}$ of $\boldsymbol{\theta}$. Therefore, for all $i = 1, \cdots, I$, we define the following function

$$
\tilde{C}_l^{(i,i)}(\boldsymbol{\theta}^{(i,i)}) \equiv C_l^{(i,i)}(\boldsymbol{\theta}).
$$

Moreover, for each pair $i, i' = 1, \cdots, I$, such that $i < i'$ the function $C_l^{(i,i')}$ only depends on the subvectors $\boldsymbol{\theta}^{(i,i')}$ and $\boldsymbol{\phi}^{(i,i')}$. Therefore, we define the following function

$$
\tilde{C}_l^{(i,i')}(\boldsymbol{\theta}^{(i,i')}, \boldsymbol{\phi}^{(i,i')}) \equiv C_l^{(i,i')}(\boldsymbol{\theta}),
$$

where $\boldsymbol{\phi}^{(i,i')} \equiv \left[ \phi_5^{(i)}, \quad \cdots, \quad \phi_K^{(i)}, \quad \phi_5^{(i')}, \quad \cdots, \quad \phi_K^{(i')} \right]^{\mathsf{T}}$.

Now, we are able to simplify the proof by introducing [4, Theorem 2.1]. This Theorem shows that under Condition **(C1)** and Condition **(C5)** the function $[\tilde{C}_{l+k}^{(i,i)}(\boldsymbol{\theta}^{(i,i)})]_{k=0,\ldots,2K-7}^{\mathsf{T}}$ is injective in $\boldsymbol{\theta}^{(i,i)}$, for all $i = 1, \cdots, I$ and all $l \in \mathbb{Z}$. Thus, $[\tilde{C}_l^{(i,i)}(\boldsymbol{\theta}^{(i,i)})]_{l \in \mathbb{Z}}^{\mathsf{T}}$ is injective. Therefore, it is sufficient to show that $[\tilde{C}_l^{(i,i')}(\boldsymbol{\theta}^{(i,i')}, \boldsymbol{\phi}^{(i,i')})]_{l \in \mathbb{Z}}^{\mathsf{T}}$ is injective, for all $i, i' = 1, \cdots, I$ such that $i < i'$. To do so, we only need to show that the following (sub)function is injective $[\tilde{C}_l^{(i,i')}(\boldsymbol{\theta}^{(i,i')}, \boldsymbol{\phi}^{(i,i')})]_{l=0,\ldots,K-4}^{\mathsf{T}}$ for all $i, i' = 1, \cdots, I$ such that $i < i'$.

Consider a pair $i, i' = 1, \cdots, I$ such that $i < i'$. Since $\boldsymbol{\phi}^{(i,i')}$ is composed of a subvector of $\boldsymbol{\theta}^{(i,i)}$ and a subvector of $\boldsymbol{\theta}^{(i',i')}$ and since $[\tilde{C}_l^{(i,i)}(\boldsymbol{\theta}^{(i,i)})]_{l \in \mathbb{Z}}^{\mathsf{T}}$ and $[\tilde{C}_l^{(i',i')}(\boldsymbol{\theta}^{(i',i')})]_{l \in \mathbb{Z}}^{\mathsf{T}}$ are injective functions, we only need to show that, for a fixed $\boldsymbol{\phi}^{(i,i')}$, the function $[\tilde{C}_l^{(i,i')}(\boldsymbol{\theta}^{(i,i')}, \boldsymbol{\phi}^{(i,i')})]_{l=0,\ldots,K-4}^{\mathsf{T}}$ is injective in $\boldsymbol{\theta}^{(i,i')}$. In other words, we want to show that the function

$$
[\tilde{C}_l^{\boldsymbol{\phi}^{(i,i')}}(\boldsymbol{\theta}^{(i,i')})]_{l=0,\ldots,K-4}^{\mathsf{T}} \equiv [\tilde{C}_l^{(i,i')}(\boldsymbol{\theta}^{(i,i')}, \boldsymbol{\phi}^{(i,i')})]_{l=0,\ldots,K-4}^{\mathsf{T}}
$$

is injective. We are going to do so by verifying Conditions **(I1)-(I4)**, which closely follows the proof of Proposition 1. Indeed, under condition **(C1)**, the CCV function is the sum of all CCV of each independent model, i.e.

$$
\tilde{C}_l^{\boldsymbol{\phi}^{(i,i')}}(\boldsymbol{\theta}^{(i,i')}) = \sigma^{(i,i')} \mathbf{1}_{l=0} + \sum_{k=5}^{K} \tilde{C}_{l,k}^{\boldsymbol{\phi}^{(i,i')}}(\boldsymbol{\theta}^{(i,i')}),
$$

where the functions $\tilde{C}_{l,k}^{\phi^{(i,i')}}(\boldsymbol{\theta}^{(i,i')})$ correspond to the ones defined in Lemma A.5, i.e.

$$\tilde{C}_{l,k}^{\phi^{(i,i')}}(\boldsymbol{\theta}^{(i,i')}) = \left(\phi_k^{(i)}\right)^{-l} \frac{z_k^{(i,i')}}{1 - \phi_k^{(i)}\,\phi_k^{(i')}}.$$

According to the expression of $\tilde{C}_l^{\phi^{(i,i')}}(\boldsymbol{\theta}^{(i,i')})$, it is trivial to see that Condition **(I2)** is satisfied, i.e.

$$|[\tilde{C}_l^{\phi^{(i,i')}}(\boldsymbol{\theta}^{(i,i')})]_{l=0,\dots,K-4}^{\mathsf{T}}|_2 \to \infty \text{ whenever } |\boldsymbol{\theta}^{(i,i')}|_2 \to \infty. \tag{4}$$

The derivative with respect to $\sigma^{(i,i')}$ is

$$a_1^{(i,i')}(l) \equiv \frac{\partial}{\partial\,\sigma^{(i,i')}} \tilde{C}_l^{\phi^{(i,i')}}(\boldsymbol{\theta}^{(i,i')})$$

$$= \begin{cases} 1 & \text{if } l = 0, \\ 0 & \text{else.} \end{cases}$$

The derivative with respect to $z_k^{(i,i')}$ is

$$a_k^{(i,i')}(l) \equiv \frac{\partial}{\partial\,z_k^{(i,i')}} \tilde{C}_l^{\phi^{(i,i')}}(\boldsymbol{\theta}^{(i,i')})$$

$$= \begin{cases} \dfrac{\left(\phi_k^{(i)}\right)^{-l}}{1 - \phi_k^{(i)}\phi_k^{(i')}} & \text{if } l < 0, \\[4mm] \dfrac{\left(\phi_k^{(i')}\right)^{-l}}{1 - \phi_k^{(i)}\phi_k^{(i')}} & \text{else.} \end{cases}$$

Since the second derivative of $[\tilde{C}_l^{\phi^{(i,i')}}(\boldsymbol{\theta}^{(i,i')})]_{l=0,\dots,K-4}^{\mathsf{T}}$ is 0, this map is in $\mathcal{C}^2(\mathbb{R}^{K-3})$ which implies Condition **(I1)**.

Let $\boldsymbol{A}$ be the Jacobian matrix of $[\tilde{C}_l^{\phi^{(i,i')}}(\boldsymbol{\theta}^{(i,i')})]_{l=0,\dots,K-4}^{\mathsf{T}}$ with respect to $\boldsymbol{\theta}^{(i,i')}$ which is given

by

$$\boldsymbol{A} = \begin{bmatrix} a_1^{(i,i')}(0) & a_5^{(i,i')}(0) & \cdots & a_K^{(i,i')}(0) \\ a_1^{(i,i')}(1) & a_5^{(i,i')}(1) & \cdots & a_K^{(i,i')}(1) \\ \vdots & \vdots & \ddots & \vdots \\ a_1^{(i,i')}(K-4) & a_5^{(i,i')}(K-4) & \cdots & a_K^{(i,i')}(K-4) \end{bmatrix}$$

$$= \begin{bmatrix} 1 & \frac{1}{1-\phi_5^{(i)}\phi_5^{(i')}} & \cdots & \frac{1}{1-\phi_K^{(i)}\phi_{K_1}^{(i')}} \\ 0 & \frac{\left(\phi_5^{(i')}\right)^{-1}}{1-\phi_5^{(i)}\phi_5^{(i')}} & \cdots & \frac{\left(\phi_K^{(i')}\right)^{-1}}{1-\phi_K^{(i)}\phi_K^{(i')}} \\ \vdots & \vdots & \ddots & \vdots \\ 0 & \frac{\left(\phi_5^{(i')}\right)^{-(K-4)}}{1-\phi_5^{(i)}\phi_5^{(i')}} & \cdots & \frac{\left(\phi_K^{(i')}\right)^{-(K-4)}}{1-\phi_K^{(i)}\phi_K^{(i')}} \end{bmatrix}.$$

Its determinant is given by

$$|\boldsymbol{A}| = \prod_{k=5}^{K} \frac{1}{1-\phi_k^{(i)}\phi_k^{(i')}} \underbrace{\begin{vmatrix} \left(\phi_5^{(i')}\right)^{-1} & \cdots & \left(\phi_K^{(i')}\right)^{-1} \\ \vdots & \ddots & \vdots \\ \left(\phi_5^{(i')}\right)^{-(K-4)} & \cdots & \left(\phi_K^{(i')}\right)^{-(K-4)} \end{vmatrix}}_{\boldsymbol{D}^{(i')}},$$

where $\boldsymbol{D}^{(i')}$ is a square Vandermonde matrix which, based on Condition **(C5)** implies that $|\boldsymbol{A}| > 0$ thereby verifying Conditions **(I3)** and **(I4)**. Therefore $[\tilde{C}_l^{\boldsymbol{\phi}^{(i,i')}}(\boldsymbol{\theta}^{(i,i')})]_{l\in\mathbb{Z}}^{\intercal}$ is injective for all pairs $1 \le i < i' \le I$. Combining this with the injectivity of $[\tilde{C}_l^{(i,i)}(\boldsymbol{\theta}^{(i,i)})]_{l\in\mathbb{Z}}^{\intercal}$ on $\boldsymbol{\theta}^{(i,i)}$, for all $1 \le i \le I$ leads to the identifiability of CCV of **(M2)**. $\square$

Using the above results, we can now prove Proposition 2.

*Proof of Proposition 2.* The proof of this proposition is a direct consequence of

1) the injectivity of the cross-covariance function $\boldsymbol{C}(\boldsymbol{\theta})$ for **(M2)** which is proved in Lemma 1,

2) the one-to-one transform between the cross-covariance function and the cross-spectral density function which is true by definition,

3) the one-to-one transform between the cross-spectral density function and the WCCV function assumed in Condition **(C6)**.

This ends the proof since a composition of injective functions is itself an injective function. $\square$

*C. Proofs of Asymptotic Properties of the MGMWM*

In order to prove the consistency of the MGMWM estimator, the following lemma states the uniform convergence of $Q_T(\boldsymbol{\theta})$ to $Q_0(\boldsymbol{\theta})$.

Lemma A.6: *Under Conditions **(C7)-(C8)**, we have that*

$$\sup_{\boldsymbol{\theta} \in \boldsymbol{\Theta}} |Q_T(\boldsymbol{\theta}) - Q_0(\boldsymbol{\theta})| \xrightarrow{p} 0.$$

*Proof of Lemma A.6.* By definition, for all $\boldsymbol{\theta} \in \boldsymbol{\Theta}$ we have

$$Q_T(\boldsymbol{\theta}) - Q_0(\boldsymbol{\theta}) = (\hat{\boldsymbol{\nu}} - \boldsymbol{\nu}(\boldsymbol{\theta}))^{\mathsf{T}} \, \boldsymbol{\Omega}_T \, (\hat{\boldsymbol{\nu}} - \boldsymbol{\nu}(\boldsymbol{\theta})) - (\boldsymbol{\nu}(\boldsymbol{\theta}_0) - \boldsymbol{\nu}(\boldsymbol{\theta}))^{\mathsf{T}} \, \boldsymbol{\Omega} \, (\boldsymbol{\nu}(\boldsymbol{\theta}_0) - \boldsymbol{\nu}(\boldsymbol{\theta})).$$

From there, a straightforward computation shows that

$$\begin{aligned}
Q_T(\boldsymbol{\theta}) - Q_0(\boldsymbol{\theta}) &= (\hat{\boldsymbol{\nu}} - \boldsymbol{\nu}(\boldsymbol{\theta}))^{\mathsf{T}} \, \boldsymbol{\Omega}_T \, (\hat{\boldsymbol{\nu}} - \boldsymbol{\nu}(\boldsymbol{\theta})) - (\boldsymbol{\nu}(\boldsymbol{\theta}_0) - \boldsymbol{\nu}(\boldsymbol{\theta}))^{\mathsf{T}} \, \boldsymbol{\Omega} \, (\boldsymbol{\nu}(\boldsymbol{\theta}_0) - \boldsymbol{\nu}(\boldsymbol{\theta})) \\
&= (\boldsymbol{\nu}(\boldsymbol{\theta}_0) - \boldsymbol{\nu}(\boldsymbol{\theta}))^{\mathsf{T}} \, (\boldsymbol{\Omega}_T - \boldsymbol{\Omega}) \, (\boldsymbol{\nu}(\boldsymbol{\theta}_0) - \boldsymbol{\nu}(\boldsymbol{\theta})) \\
&\quad + (\hat{\boldsymbol{\nu}} - \boldsymbol{\nu}(\boldsymbol{\theta}_0))^{\mathsf{T}} \, \boldsymbol{\Omega}_T \, (\hat{\boldsymbol{\nu}} - \boldsymbol{\nu}(\boldsymbol{\theta}_0)) \\
&\quad - 2 \, (\boldsymbol{\nu}(\boldsymbol{\theta}_0) - \boldsymbol{\nu}(\boldsymbol{\theta}))^{\mathsf{T}} \, \boldsymbol{\Omega}_T \, (\boldsymbol{\nu}(\boldsymbol{\theta}_0) - \hat{\boldsymbol{\nu}}).
\end{aligned}$$

Therefore, using the triangular inequality, we have

$$\begin{aligned}
\sup_{\boldsymbol{\theta} \in \boldsymbol{\Theta}} |Q_T(\boldsymbol{\theta}) - Q_0(\boldsymbol{\theta})| \leq &\sup_{\boldsymbol{\theta} \in \boldsymbol{\Theta}} \left| (\boldsymbol{\nu}(\boldsymbol{\theta}_0) - \boldsymbol{\nu}(\boldsymbol{\theta}))^{\mathsf{T}} \, (\boldsymbol{\Omega}_T - \boldsymbol{\Omega}) \, (\boldsymbol{\nu}(\boldsymbol{\theta}_0) - \boldsymbol{\nu}(\boldsymbol{\theta})) \right| \\
&+ \left| (\hat{\boldsymbol{\nu}} - \boldsymbol{\nu}(\boldsymbol{\theta}_0))^{\mathsf{T}} \, \boldsymbol{\Omega}_T \, (\hat{\boldsymbol{\nu}} - \boldsymbol{\nu}(\boldsymbol{\theta}_0)) \right| \\
&+ 2 \sup_{\boldsymbol{\theta} \in \boldsymbol{\Theta}} \left| (\boldsymbol{\nu}(\boldsymbol{\theta}_0) - \boldsymbol{\nu}(\boldsymbol{\theta}))^{\mathsf{T}} \, \boldsymbol{\Omega}_T \, (\hat{\boldsymbol{\nu}} - \boldsymbol{\nu}(\boldsymbol{\theta}_0)) \right|.
\end{aligned}$$

We are going to analyse each term of the right-hand side of the latter inequality. For the first term, we have

$$\begin{aligned}
&\sup_{\boldsymbol{\theta} \in \boldsymbol{\Theta}} \left| (\boldsymbol{\nu}(\boldsymbol{\theta}_0) - \boldsymbol{\nu}(\boldsymbol{\theta}))^{\mathsf{T}} \, (\boldsymbol{\Omega}_T - \boldsymbol{\Omega}) \, (\boldsymbol{\nu}(\boldsymbol{\theta}_0) - \boldsymbol{\nu}(\boldsymbol{\theta})) \right| \\
&\leq \sup_{\boldsymbol{\theta} \in \boldsymbol{\Theta}} |\boldsymbol{\nu}(\boldsymbol{\theta}_0) - \boldsymbol{\nu}(\boldsymbol{\theta})|_2^2 ||\boldsymbol{\Omega}_T - \boldsymbol{\Omega}||_S \\
&= \sup_{\boldsymbol{\theta} \in \boldsymbol{\Theta}} \sum_{i=1}^{I} \sum_{i'=i}^{I} \sum_{j=1}^{J} \left( \gamma_j^{(i,i')}(\boldsymbol{\theta}_0) - \gamma_j^{(i,i')}(\boldsymbol{\theta}) \right)^2 ||\boldsymbol{\Omega}_T - \boldsymbol{\Omega}||_S.
\end{aligned} \tag{5}$$

Condition **(C7)** implies that for all $i, i' = 1, \ldots, I$ and all $j = 1, \ldots, J$, there exists $B_j^{(i,i')} \in \mathbb{R}$

such that for all $\boldsymbol{\theta}, \boldsymbol{\theta}_0 \in \boldsymbol{\Theta}$, we have

$$\left(\gamma_j^{(i,i')}(\boldsymbol{\theta}_0) - \gamma_j^{(i,i')}(\boldsymbol{\theta})\right)^2 \leq B_j^{(i,i')}.$$

Therefore using conditions **(C7)** and **(C8)**, Eq. (5) yields

$$\sup_{\boldsymbol{\theta} \in \boldsymbol{\Theta}} \left| \left(\boldsymbol{\nu}(\boldsymbol{\theta}_0) - \boldsymbol{\nu}(\boldsymbol{\theta})\right)^{\mathsf{T}} \left(\boldsymbol{\Omega}_T - \boldsymbol{\Omega}\right) \left(\boldsymbol{\nu}(\boldsymbol{\theta}_0) - \boldsymbol{\nu}(\boldsymbol{\theta})\right) \right|$$

$$\leq \sup_{\boldsymbol{\theta} \in \boldsymbol{\Theta}} \sum_{i=1}^{I} \sum_{i'=i}^{I} \sum_{j=1}^{J} \left(\gamma_j^{(i,i')}(\boldsymbol{\theta}_0) - \gamma_j^{(i,i')}(\boldsymbol{\theta})\right)^2 ||\boldsymbol{\Omega}_T - \boldsymbol{\Omega}||_S$$

$$\leq \sum_{i=1}^{I} \sum_{i'=i}^{I} \sum_{j=1}^{J} B_j^{(i,i')} ||\boldsymbol{\Omega}_T - \boldsymbol{\Omega}||_S \xrightarrow{p} 0.$$

For the second term, let $\lambda_T \equiv ||\boldsymbol{\Omega}_T||_S$. Then by Condition **(C8)** and Theorem 1, we have

$$\left| \left(\hat{\boldsymbol{\nu}} - \boldsymbol{\nu}(\boldsymbol{\theta}_0)\right)^{\mathsf{T}} \boldsymbol{\Omega}_T \left(\hat{\boldsymbol{\nu}} - \boldsymbol{\nu}(\boldsymbol{\theta}_0)\right) \right| \leq \lambda_T |\hat{\boldsymbol{\nu}} - \boldsymbol{\nu}(\boldsymbol{\theta}_0)|_2^2 \xrightarrow{p} 0.$$

For the last term, applying Theorem 1 again, we have

$$2 \sup_{\boldsymbol{\theta} \in \boldsymbol{\Theta}} \left| \left(\boldsymbol{\nu}(\boldsymbol{\theta}_0) - \boldsymbol{\nu}(\boldsymbol{\theta})\right)^{\mathsf{T}} \boldsymbol{\Omega}_T \left(\hat{\boldsymbol{\nu}} - \boldsymbol{\nu}(\boldsymbol{\theta}_0)\right) \right|$$

$$\leq 2 \sup_{\boldsymbol{\theta} \in \boldsymbol{\Theta}} \left( |\boldsymbol{\nu}(\boldsymbol{\theta}_0) - \boldsymbol{\nu}(\boldsymbol{\theta})|_2 \, ||\boldsymbol{\Omega}_T||_S \, |\hat{\boldsymbol{\nu}} - \boldsymbol{\nu}(\boldsymbol{\theta}_0)|_2 \right)$$

$$\leq 2 \left( \frac{(I+1)IJ}{2} \max_{i,i',j} B_j^{(i,i')} \right)^{1/2} \lambda_T \, o_p(1) \xrightarrow{p} 0.$$

We are now able to conclude the proof with the following computation,

$$\sup_{\boldsymbol{\theta} \in \boldsymbol{\Theta}} |Q_T(\boldsymbol{\theta}) - Q_0(\boldsymbol{\theta})|$$

$$\leq \underbrace{\sup_{\boldsymbol{\theta} \in \boldsymbol{\Theta}} \left| \left(\boldsymbol{\nu}(\boldsymbol{\theta}_0) - \boldsymbol{\nu}(\boldsymbol{\theta})\right)^{\mathsf{T}} \left(\boldsymbol{\Omega}_T - \boldsymbol{\Omega}\right) \left(\boldsymbol{\nu}(\boldsymbol{\theta}_0) - \boldsymbol{\nu}(\boldsymbol{\theta})\right) \right|}_{\xrightarrow{p} 0}$$

$$+ \underbrace{\left| \left(\hat{\boldsymbol{\nu}} - \boldsymbol{\nu}(\boldsymbol{\theta}_0)\right)^{\mathsf{T}} \boldsymbol{\Omega}_T \left(\hat{\boldsymbol{\nu}} - \boldsymbol{\nu}(\boldsymbol{\theta}_0)\right) \right|}_{\xrightarrow{p} 0}$$

$$+ \underbrace{2 \sup_{\boldsymbol{\theta} \in \boldsymbol{\Theta}} \left| \left(\boldsymbol{\nu}(\boldsymbol{\theta}_0) - \boldsymbol{\nu}(\boldsymbol{\theta})\right)^{\mathsf{T}} \boldsymbol{\Omega}_T \left(\hat{\boldsymbol{\nu}} - \boldsymbol{\nu}(\boldsymbol{\theta}_0)\right) \right|}_{\xrightarrow{p} 0} \xrightarrow{p} 0.$$

$\square$

*Proof of Theorem* **2**. Based on Condition **(C9)** and the non-singularity of $\Omega$, $Q_0(\boldsymbol{\theta})$ has a unique minimum at $\boldsymbol{\theta}_0$. Therefore, by Conditions **(C7)**-**(C10)** and using Lemma **A.6**, [**5**, Theorem 2.1] can be applied implying the consistency of $\hat{\boldsymbol{\theta}}$.

$\square$

*Proof of Theorem* **3**. By Lemma **A.3**, **A.4** and **A.5**, $\boldsymbol{\nu}(\boldsymbol{\theta})$ is continuously differentiable. Therefore, by definition of $\hat{\boldsymbol{\theta}}$, we have

$$\left. \frac{\partial Q_T(\boldsymbol{\theta})}{\partial \boldsymbol{\theta}} \right|_{\boldsymbol{\theta}=\hat{\boldsymbol{\theta}}} = \mathbf{0}_{p \times 1}$$

$$\Longleftrightarrow \left. \frac{\partial}{\partial \boldsymbol{\theta}} \left[ (\hat{\boldsymbol{\nu}} - \boldsymbol{\nu}(\boldsymbol{\theta}))^{\intercal} \, \boldsymbol{\Omega}_T \, (\hat{\boldsymbol{\nu}} - \boldsymbol{\nu}(\boldsymbol{\theta})) \right] \right|_{\boldsymbol{\theta}=\hat{\boldsymbol{\theta}}} = \mathbf{0}_{p \times 1}$$

which, up to a constant, yields

$$\underbrace{\left( \frac{\partial}{\partial \boldsymbol{\theta}} \left( \hat{\boldsymbol{\nu}} - \boldsymbol{\nu}(\boldsymbol{\theta}) \right)^{\intercal} \big|_{\boldsymbol{\theta}=\hat{\boldsymbol{\theta}}} \right)}_{\boldsymbol{B}(\hat{\boldsymbol{\theta}})} \boldsymbol{\Omega}_T \left( \hat{\boldsymbol{\nu}} - \boldsymbol{\nu}(\hat{\boldsymbol{\theta}}) \right) = \mathbf{0}_{p \times 1}. \tag{6}$$

The multivariate mean value theorem ensures that, based on Condition **(C11)**, there exists a matrix $A(\hat{\boldsymbol{\theta}}, \boldsymbol{\theta}_0)$ that can be used to expand $\hat{\boldsymbol{\nu}} - \boldsymbol{\nu}(\hat{\boldsymbol{\theta}})$ around $\boldsymbol{\theta}_0$ in the following way

$$\hat{\boldsymbol{\nu}} - \boldsymbol{\nu}(\hat{\boldsymbol{\theta}}) = \hat{\boldsymbol{\nu}} - \boldsymbol{\nu}(\boldsymbol{\theta}_0) + \boldsymbol{A}(\hat{\boldsymbol{\theta}}, \boldsymbol{\theta}_0) \left( \hat{\boldsymbol{\theta}} - \boldsymbol{\theta}_0 \right). \tag{7}$$

Since $\hat{\boldsymbol{\nu}} \xrightarrow{p} \boldsymbol{\nu}(\boldsymbol{\theta}_0)$ and $\hat{\boldsymbol{\theta}} \xrightarrow{p} \boldsymbol{\theta}_0$, the multivariate mean value theorem also guarantees that the matrix $\boldsymbol{A}(\hat{\boldsymbol{\theta}}, \boldsymbol{\theta}_0)$ has the following property

$$\boldsymbol{A}(\hat{\boldsymbol{\theta}}, \boldsymbol{\theta}_0) \xrightarrow{p} \left. \frac{\partial}{\partial \boldsymbol{\theta}^{\intercal}} \left( \boldsymbol{\nu}(\boldsymbol{\theta}_0) - \boldsymbol{\nu}(\boldsymbol{\theta}) \right) \right|_{\boldsymbol{\theta}=\boldsymbol{\theta}_0} = \left. \frac{\partial}{\partial \boldsymbol{\theta}^{\intercal}} \boldsymbol{\nu}(\boldsymbol{\theta}) \right|_{\boldsymbol{\theta}=\boldsymbol{\theta}_0}, \tag{8}$$

given that $\partial/\partial\boldsymbol{\theta}\,\boldsymbol{\nu}(\boldsymbol{\theta}^{\intercal})$ is continuous (based also on Lemmas **A.3**, **A.4** and **A.5**). We now use the notation $\boldsymbol{A}(\boldsymbol{\theta}_0) \equiv \partial/\partial\boldsymbol{\theta}^{\intercal}\,\boldsymbol{\nu}(\boldsymbol{\theta})|_{\boldsymbol{\theta}=\boldsymbol{\theta}_0}$ and note that, plugging (7) in the third factor of (6), multiplying by $\sqrt{T}$ and using Condition **(C12)**, we have that

$$\sqrt{T} \left( \hat{\boldsymbol{\theta}} - \boldsymbol{\theta}_0 \right) = - \left[ \boldsymbol{B}(\hat{\boldsymbol{\theta}}) \, \boldsymbol{\Omega}_T \, \boldsymbol{A}(\hat{\boldsymbol{\theta}}, \boldsymbol{\theta}_0) \right]^{-1} \boldsymbol{B}(\hat{\boldsymbol{\theta}}) \, \boldsymbol{\Omega}_T \, \sqrt{T} \left( \hat{\boldsymbol{\nu}} - \boldsymbol{\nu}(\boldsymbol{\theta}_0) \right). \tag{9}$$

Knowing that $\boldsymbol{B}(\hat{\boldsymbol{\theta}}) \xrightarrow{p} \boldsymbol{A}(\boldsymbol{\theta}_0)^{\intercal}$ by the continuous mapping theorem, by Slutsky's theorem we have that

$$\left[ \boldsymbol{B}(\hat{\boldsymbol{\theta}}) \, \boldsymbol{\Omega}_T \, \boldsymbol{A}(\hat{\boldsymbol{\theta}}, \boldsymbol{\theta}_0) \right]^{-1} \boldsymbol{B}(\hat{\boldsymbol{\theta}}) \, \boldsymbol{\Omega}_T \xrightarrow{p} \left[ \boldsymbol{A}(\boldsymbol{\theta}_0)^{\intercal} \boldsymbol{\Omega} \boldsymbol{A}(\boldsymbol{\theta}_0) \right]^{-1} \boldsymbol{A}(\boldsymbol{\theta}_0)^{\intercal} \boldsymbol{\Omega}$$

By again using Slutsky's theorem as well Theorem 1 we have that (9) has the following asymptotic distribution

$$\sqrt{T}\left(\hat{\boldsymbol{\theta}} - \boldsymbol{\theta}_0\right) \xrightarrow{\mathcal{D}} \mathcal{N}(\boldsymbol{0}, \boldsymbol{\Xi}),$$

where, denoting $\boldsymbol{G}(\boldsymbol{\theta}_0) \equiv [\boldsymbol{A}(\boldsymbol{\theta}_0)^\intercal \boldsymbol{\Omega} \boldsymbol{A}(\boldsymbol{\theta}_0)]^{-1}$, the asymptotic covariance matrix $\boldsymbol{\Xi}$ is given by

$$\boldsymbol{\Xi} \equiv \boldsymbol{G}(\boldsymbol{\theta}_0)\boldsymbol{A}(\boldsymbol{\theta}_0)^\intercal \boldsymbol{\Omega} \mathbb{E}(\boldsymbol{D}_0 \boldsymbol{D}_0^\intercal) \boldsymbol{\Omega} \boldsymbol{A}(\boldsymbol{\theta}_0) \boldsymbol{G}(\boldsymbol{\theta}_0).$$

and $\boldsymbol{D}_0$ is given in Theorem 1. □

### D. Simulations

In this section we give some additional simulation results. Firstly, the boxplots in Fig. A.1 represent the results of the simulations in the main paper. With respect to the summary information in the MSE provided in Fig. 1, it is clear from these boxplots how the MGMWM is a much more statistically efficient estimator compared to the GLS(2) (and GLS(3)) thereby explaining the significant difference between their MSEs.

Secondly, we present a second simulation study which is made under a more general and flexible setting which cannot be performed using the approach in [6]. More specifically, we simulate an array of two univariate processes, where the first process is made by the sum of **(T1)** model (WN), a **(T5)** model (AR(1)) and an ARMA(1,1) model. The latter is not explicitly listed among the possible models considered in this paper but can nevertheless be represented as a compact reparametrized sum of a WN and AR(1) model. The second simulated processes is made by the sum of **(T1)** model (WN), a **(T5)** model (AR(1)) and a **(T2)** model (RW). For both simulated processes, the WN and the AR(1) models are correlated while the ARMA component of the first process is independent of the RW component of the second process. Moreover, within each process, all the components are independent. The simulation setting and results are represented through boxplots shown in Fig. A.2.

Although it currently cannot be compared to any other estimator, it is clear that the MGMWM can be used to estimate more complex model- and interdependence-settings than those considered in the main paper.

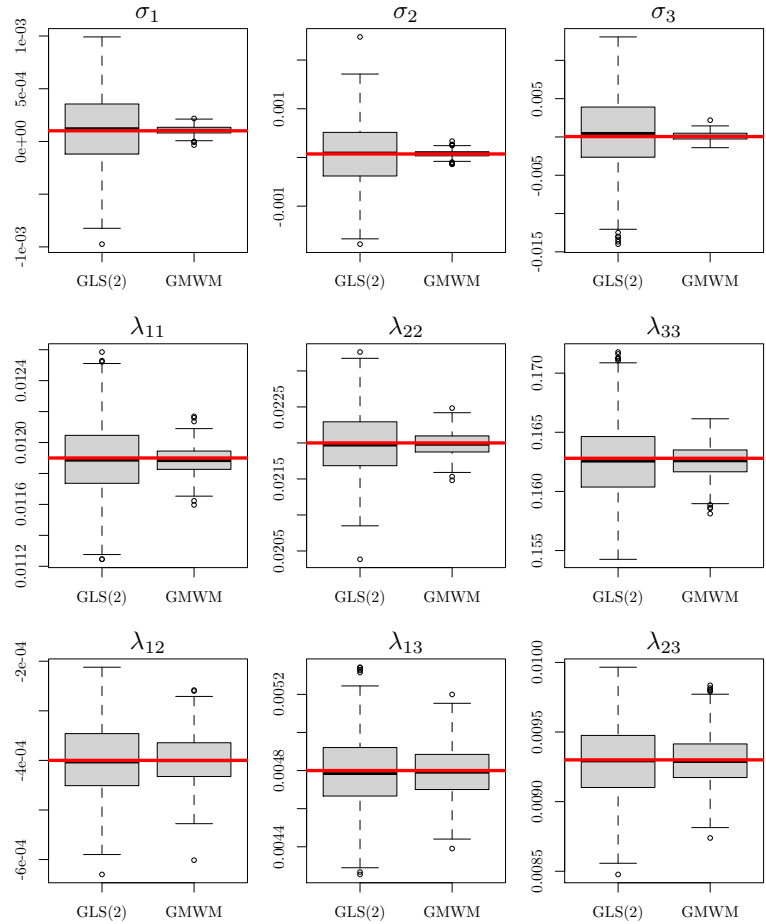

Fig. A.1. Boxplots for the parameters of a multivariate model (array of three sensors) made by the sum of multivariate WN and RW models ($\sigma_i$ representing the $i^{th}$ parameter for the WN model and $\lambda_{ii'}$ representing a component of the covariance matrix defining the multivariate RW model.

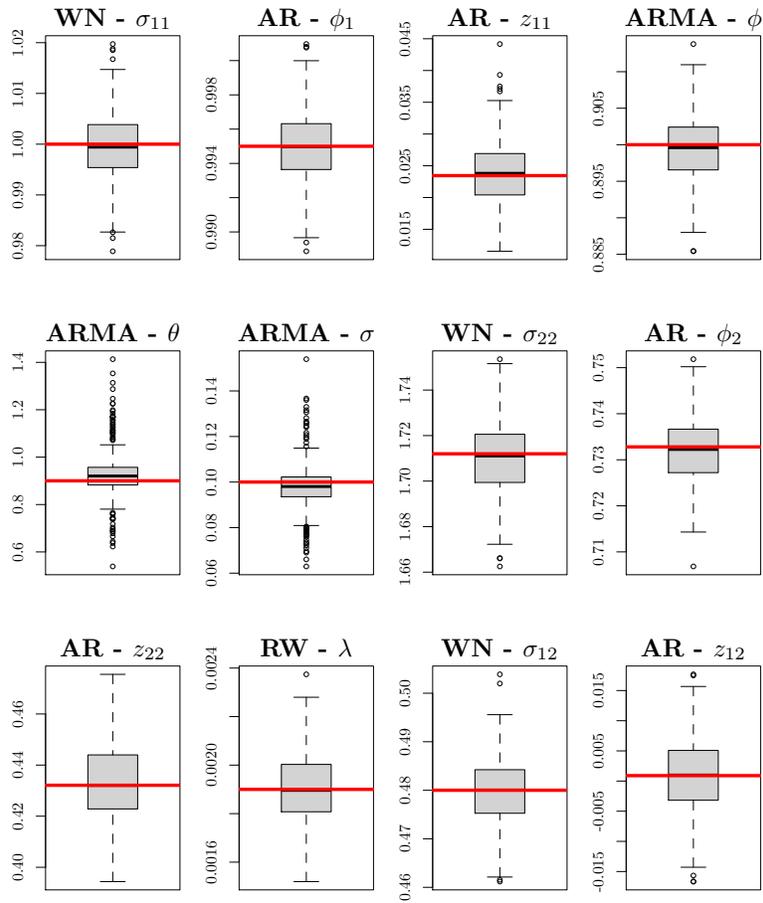

Fig. A.2. Boxplots representing the empirical distribution of MGMWM estimations for an array of two sensors where the first signal is generated by a multivariate WN + AR(1) + ARMA(1,1) model while the second signal is generated by a WN + AR(1) + RW model where the WN and AR(1) components of the two signals are correlated. The boxplots are based on 500 replications and sample size $T = 10^6$.

}

\end{document}